\documentclass[11pt,a4paper]{article}

\usepackage[margin=1in]{geometry}
\usepackage{amsmath,amssymb,amsthm}
\usepackage{graphics}
\usepackage{hyperref}
\usepackage{physics}
\usepackage{cite}
\usepackage{todonotes}
\usepackage{comment}
\usepackage{authblk}
\usepackage{optidef}
\usepackage{graphicx}
\usepackage{subcaption}
\usepackage{booktabs}
\title{Unconstrained Binary Models of the Travelling Salesman Problem Variants for Quantum Optimization}
\author[1]{{\"Ozlem Salehi}\thanks{Corresponding author: osalehi@iitis.pl}}
\author[1]{Adam Glos}
\author[1]{Jarosław Adam Miszczak}
\affil[1]{Institute of Theoretical and Applied Informatics, Polish Academy of Sciences, Bałtycka~5, 44-100 Gliwice, Poland}
\date{}

\newcommand{\ZZ}{\mathbb Z}
\newcommand{\RR}{\mathbb R}

\newcommand{\edges}{\vec E}

\begin{document}
	\maketitle
	\begin{abstract}
		
Quantum computing is offering a novel perspective for solving combinatorial optimization problems. To fully explore the possibilities offered by quantum computers, the problems need to be formulated as unconstrained binary models, taking into account limitation and advantages of quantum devices. In this work, we provide a detailed analysis of the Travelling Salesman Problem with Time Windows (TSPTW) in the context of solving it on a quantum computer. We introduce quadratic unconstrained binary optimization and higher order binary optimization formulations of this problem. We demonstrate the advantages of edge-based and node-based formulations of the TSPTW problem. Additionally, we investigate the experimental realization of the presented methods on a quantum annealing device. The provided results pave the path for utilizing quantum computer for a variety of real-world task which can be cast in the form of Travelling Salesman Problem with Time Windows problem.
\end{abstract}

\section{Introduction}

The well-known Travelling Salesman Problem (TSP) and its variants have been widely studied and posses a rich literature~\cite{tsp2002,tsp2011}. Among its variants, Travelling Salesman Problem with Time Windows (TSPTW) looks for a tour with the minimum cost where each city is visited within an associated time frame, namely between its earliest start time and its due time. Such formulated time-constrained problems have many industrial applications in a variety of fields, including logistics, transportation systems, and manufacturing \cite{desrosiers1995time}. Furthermore, investigation of TSPTW allows a deeper understanding of other related and more complicated problems like the vehicle routing problem with time windows \cite{el2010vehicle} and its variants. It is proven that the TSPTW problem is NP-Hard, and even finding a feasible solution is proven to be NP-Complete~\cite{savelsbergh1985local}.

It is not surprising that TSP and its variants provide a paradigmatic benchmark for the emerging technology of quantum computing, which opens up an alternative perspective for solving computationally hard problems. Quantum algorithms developed by Shor and Grover, with provable speedups compared to the best known classical algorithms, are not suitable for the noisy intermediate-scale quantum (NISQ) era \cite{preskill2018quantum}. At the same time, there have been promising attempts to solve optimization problems, including TSP, using current quantum technology, the most prominent ones being the Variational Quantum Eigensolver (VQE) \cite{peruzzo2014variational}, Quantum Approximate Optimization Algorithm (QAOA) \cite{farhi2014quantum}, and quantum annealing (QA) \cite{apolloni1989quantum, kadowaki1998quantum}. In each case one needs to represent the optimization problem in the form of  unconstrained binary model.

In particular, quantum annealing relies on the quantum adiabatic theorem \cite{farhi2000quantum}. An initial Hamiltonian is picked whose ground state is easy to prepare, and the system is evolved by applying a time-dependent Hamiltonian. The time-dependent Hamiltonian gradually brings in the problem Hamiltonian, whose ground state encodes the solution to the optimization problem of interest. Quantum adiabatic theorem guarantees that the final state of the system is close to the ground state of the problem Hamiltonian  for a sufficiently long evolution \cite{childs2001robustness}. Hence, finding the solution to the optimization problem can be reduced to the problem of finding the ground state of the problem Hamiltonian. Theoretically, the performance of quantum annealing depends on the minimum spectral gap encountered during the process. Whether quantum annealing provides speedup against the classical algorithms remains controversial \cite{McGeoch2020theory,hauke2020perspectives}. How to detect quantum speedup is a question of interest on its own \cite{ronnow2014defining}, and various metrics have been proposed.

Quantum annealing has attracted significant attention since it is realizable in the commercially available D-Wave machines \cite{johnson2011quantum}. In order to solve a problem using QA, one can formulate the problem as a quadratic unconstrained binary optimization problem (QUBO) which is then easily recast into the problem Hamiltonian. The only QUBO formulation proposed in the literature for the TSPTW problem is due to Papalitsas et al. \cite{papalitsas2019qubo}. However, the formulation is flawed as it allows some infeasible solutions. Furthermore, the authors assume that the earliest start time is equal to 0 for all cities, which limits the possible applications.

In this paper, we provide three different formulations for the TSPTW problem. The first formulation extends the formulation proposed in \cite{papalitsas2019qubo} by taking into account both the earliest start times and the due times for each city. In the second formulation, we provide an alternative higher-order binary model which is more space-efficient. Finally, we present an alternative QUBO formulation based on the integer linear programming formulation given in \cite{kara2013new}. First two formulations allow more than one assignment of binary variables to encode the optimal route without any penalty. All proposed models can be easily modified to obtain formulations for other variants of TSP like Makespan Problem with Time Windows, in which the total tour duration is minimized. To investigate the efficiency of the edge-based and ILP formulations, we provide some experimental results obtained by running small instances of the problem on the D-Wave Advantage.

This paper is organized into the following sections. We start with introducing the problem and the necessary concepts in Section \ref{sec:back}. In Section \ref{sec:formulation} we present our formulations for the TSPTW problem. Section \ref{sec:results} contains experimental results from the D-Wave quantum annealing device. We discuss the formulations and the results in Section \ref{sec:discussion} and we conclude by Section \ref{sec:conclusion}.

%%%%%%%%%%%%%%%%%%%%%%%%%%%%%%%%%%%%%%%%%%%%%%%%%%%%%%%%%%%%%%%%%%%%%%%%%%%%%%%%
\section{Background}\label{sec:back} 
%%%%%%%%%%%%%%%%%%%%%%%%%%%%%%%%%%%%%%%%%%%%%%%%%%%%%%%%%%%%%%%%%%%%%%%%%%%%%%%%

In this section, we provide background information concerning quantum annealing and related concepts. We also discuss recent results concerning Travelling Salesman Problem with Time Windows.

%%%%%%%%%%%%%%%%%%%%%%%%%%%%%%%%%%%%%%%%%%%%%%%%%%%%%%%%%%%%%%%%%%%%%%%%%%%%%%%%
\subsection{Simulated and quantum annealing}  
%%%%%%%%%%%%%%%%%%%%%%%%%%%%%%%%%%%%%%%%%%%%%%%%%%%%%%%%%%%%%%%%%%%%%%%%%%%%%%%%

Consider a minimization problem in which the aim is to find the global minimum of a cost function defined over a discrete set, whose elements are called the solutions. Let's model the solution space as a graph where the nodes are the solutions, and the edges are defined by a neighbourhood rule. Proposed by Kirkpatrick et al. \cite{kirkpatrick1983optimization}, \textbf{simulated annealing} (SA) can be regarded as a random walk on the search space, whose steps are parametrized by a temperature parameter called $T$. At each iteration, a random step is taken, and the step is accepted if it has a lower cost. If the new step has a higher cost, then it is accepted with a probability determined by the temperature $ T $ and the difference between the existing and the new costs. At higher temperatures, the transition between the states occurs more frequently. According to a cooling schedule, at each iteration, the parameter $ T $ is decremented, and the optimal solution is found with the help of the thermal fluctuations.      

\textbf{Quantum annealing} (QA) is a quantum mechanical heuristic method for solving optimization problems relying on the quantum adiabatic computation. Note that quantum annealing is indeed a physical process taking place in an analog quantum device, whereas simulated annealing is an analogy of a physical procedure. The idea is independently introduced by many authors, including \cite{apolloni1989quantum, kadowaki1998quantum}. As opposed to simulated annealing, in quantum annealing, quantum fluctuations are used instead of thermal fluctuations.

In the framework of QA, the system is initialized to the ground state of the transverse field Hamiltonian $H_D$ and a problem Hamiltonian $H_F$ is designed so that its ground state encodes the solution. The overall Hamiltonian takes the form
\[ 
H(t) = A(t) H_D + B(t)H_F,
\]
where $ A(t) $ is decreased gradually from the initial value $ A(0)=1 $ to $ A(\tau) =0$ while $ B(t) $ is gradually increased from $ B(0)=0 $ to $ B(\tau)=1 $, $ \tau $ being the computation time, so that at the end of the annealing procedure $ H(\tau) = H_F $. Then by quantum adiabatic theorem, the final ground state gives us a solution which is close to the optimal. Instead of climbing over the local minima as in SA, quantum tunneling is used to escape local minima \cite{mcgeoch2014adiabatic}. We refer readers to \cite{mcgeoch2014adiabatic, hauke2020perspectives} to read more details on the topic. 

%%%%%%%%%%%%%%%%%%%%%%%%%%%%%%%%%%%%%%%%%%%%%%%%%%%%%%%%%%%%%%%%%%%%%%%%%%%%%%%%
\subsection{Ising model and D-Wave}
%%%%%%%%%%%%%%%%%%%%%%%%%%%%%%%%%%%%%%%%%%%%%%%%%%%%%%%%%%%%%%%%%%%%%%%%%%%%%%%%

Consider $n$ particles where each particle can be either in state $-1$ or $+1$  called the spin. An assignment of $-1$s and 1s to the spins is known as the spin configuration. The \textbf{Ising model} is a mathematical model for ferromagnetism used in statistical mechanics to analyze the properties of spin configurations. The interaction force or the coupling strength between the particles is denoted by $ J_{ij} $, and an external force $h_i$ called the qubit bias is applied on each particle. The energy of a configuration is given by
\[ 
H(s) = \sum_{i}h_i s_i + \sum_{i<j} J_{ij}s_i s_j,
\] 
where $ s_i \in \{-1,1\} $. One should note that the problem of finding the spin configuration which minimizes $H(s)$ is NP-hard in general. 

D-Wave QPU is a collection of particles arranged on a special architecture (Chimera or Pegasus), and the Ising energy minimization problem is natively solved using quantum annealing. The initial and the problem Hamiltonians take the form
\[ 
H_D = \sum_i \sigma_i^x, ~~~~H_F = \sum_i h_i \sigma_i^z + \sum_{i<j} J_{ij} \sigma_i^z \sigma_j^z,
\]  
where $ \sigma_i^x $ and $ \sigma_i^z $ denote the Pauli-$ x $ and Pauli-$ z $ operators acting on $ i$-th qubit respectively. Hence, one needs to design the Hamiltonian $ H_F $ whose ground state encodes the optimal solution to the minimization problem that is aimed to be solved. 

Note that not all couplings are available on the hardware, and therefore a process called minor embedding is needed to map the logical qubits to the physical ones. Furthermore, there are specific ranges for $ h_i$ and $ J_{ij} $, so that the coupling and the qubit bias require scaling.

%%%%%%%%%%%%%%%%%%%%%%%%%%%%%%%%%%%%%%%%%%%%%%%%%%%%%%%%%%%%%%%%%%%%%%%%%%%%%%%%
\subsection{Quadratic Unconstrained Binary Optimization}
%%%%%%%%%%%%%%%%%%%%%%%%%%%%%%%%%%%%%%%%%%%%%%%%%%%%%%%%%%%%%%%%%%%%%%%%%%%%%%%%

One may find it more natural to express an optimization problem using binary variables instead of spin variables. Formally, objective function for the \textbf{Quadratic Unconstrained Binary Optimization} problem is defined as:
\begin{align*}
H(x) =  \sum_{i\leq j} x_i Q_{ij} x_j,
\end{align*}
where $ x $ is a binary vector and $ Q $ is a real square upper triangular matrix. The correspondence between the QUBO and the Ising Model can be observed by the change of variable $ x_i = \frac{1-s_i}{2} $. Noting that $ x_i^2=x_i $ for any $ i $ since $ x_i \in \{0,1\} $, the diagonals of the matrix $ Q $ are the linear coefficients and the off-diagonals are the quadratic coefficients. In \cite{lucas2014ising}, a list of well-known problems and their QUBO formulations are presented.

One can propose a generalization of QUBO in which the objective function is a general polynomial in bits. If the objective function involves monomials with  $ k \geq 3$ variables, then it is called a high order binary optimization (HOBO) problem. Note that sometimes such problem
is also referred to as PUBO, polynomial unconstrained binary
optimization problem~\cite{perdomo2019readiness}. It is always possible to obtain an equivalent QUBO formulation by quadratization, which is explained in more detail in the Appendix.

%%%%%%%%%%%%%%%%%%%%%%%%%%%%%%%%%%%%%%%%%%%%%%%%%%%%%%%%%%%%%%%%%%%%%%%%%%%%%%%%
\subsection{Integer Linear Programming}
%%%%%%%%%%%%%%%%%%%%%%%%%%%%%%%%%%%%%%%%%%%%%%%%%%%%%%%%%%%%%%%%%%%%%%%%%%%%%%%%

In \textbf{integer linear programming} (ILP), the problems are formulated through some set of linear constraints over integer variables and a linear objective function to be minimized. An ILP problem is formally defined as 
\begin{alignat*}{3}
&\text{minimize} \hspace{1em}&& \sum_{i} c_i y_i \\
&\text{subject to} \hspace{1em }&& \sum_{j} a_{ij}y_j \leq b_i, \hspace{1em} i=1,\dots,m \\
&{} &&y_i \geq 0, y_i \in \ZZ   
\end{alignat*}
where $a_{ij}\in \RR$, $b_j\in \RR$, $ c_i \in \RR $. ILP problem is known to be NP-Complete.

Integer quadratic program (IQP) is defined analogously with a quadratic objective function and a set of linear constraints. Both ILP and IQP problems can be expressed as QUBO problems which will be discussed next.

%%%%%%%%%%%%%%%%%%%%%%%%%%%%%%%%%%%%%%%%%%%%%%%%%%%%%%%%%%%%%%%%%%%%%%%%%%%%%%%%
\subsection{Transformation into binary problems} \label{sec:qubo}
%%%%%%%%%%%%%%%%%%%%%%%%%%%%%%%%%%%%%%%%%%%%%%%%%%%%%%%%%%%%%%%%%%%%%%%%%%%%%%%%

Direct preparation of QUBO formulation may not always be convenient. One may first define some constraints and use integer variables while formulating an optimization problem. Below we present the procedures for transforming linear inequalities into equality constraints, mapping integer variables to binary ones, and the penalty method for removing the constraints. 

Suppose we have integer variables $y_1,\dots,y_k$ such that $\underline y_i \leq y_i \leq \overline y_i$ where $\underline y_i,\overline y_i\in \ZZ$ are some constants bounding $ y_i $. Since in the paper we will use integer variables only, we will write  $y\in \{\underline{y}_i,\ldots, \overline{y}_i\}$ instead of $\underline y_i \leq y_i \leq \overline y_i$ . Also suppose that $f(y_1,\dots,y_k)$ is our objective function to be minimized. 

Let us start with the penalty method for removing the constraints. Given a linear equality constraint of the form
\begin{equation}
	\sum_{i=1}^ka_i y_i = b 
\end{equation}
where $a_i,b\in \RR$, the transformation procedure simply transforms objective function $f$ into
\begin{equation}
f(y_1,\dots,y_k) + P \left(\sum_{i=1}^ka_i y_i - b\right)^2.
\end{equation}
Note that the new function is equal to $f$ if and only if variables $y_i$ satisfy the equality. Constant $P\in \RR_{>0}$ is the penalty constant that has to be adjusted.

Linear inequality constraints have to be first transformed to equality constraints first through so-called slack variables. Suppose we have an inequality constraint of the form 
\begin{equation}
\sum_{i=1}^ka_i y_i \leq b.
\end{equation}
Then by adding slack variables $\xi$, we obtain $\sum_{i=1}^ka_i y_i + \xi = b$, and move it to the objective function according to previously described procedure. Note that $\xi$ has to be optimized by the optimization procedure as well. Taking into account the both side of the inequality, we can bound the slack variable $ \xi $ as follows:
\begin{equation}
0 \leq \xi \leq - \left (\sum_{i=1}^k \min\{ a_{i}\underline y_i, a_{i}\overline y_i\} -b \right ).
\end{equation}

Now suppose we have a function $f(y_1,\dots,y_k)$, where $y_1,\dots,y_k$ are integer variables. In order to transform $f$ into pseudo-Boolean function, it is enough to replace each occurrence of an integer variable $y$ with 
\begin{equation}
E_{\underline y}^{\overline{y}}(y) = \underline y+\sum_{i=0}^{k_y-2}2^i x_{y,i}+\bigl (\overline{y}-\sum_{i=0}^{k_y-2}2^i\bigr)x_{y,k_y-1},
\end{equation}
where $k_{y} = \lceil \log_2(\overline y-\underline y+1)\rceil $, and $x_{y,i}$ are the newly introduced binary variables to be optimized. Note that $f$ and $f(E_{\underline y_1}^{\overline{y_1}}(y_1),\dots,E_{\underline y_k}^{\overline{y_k}}(y_k))$ are polynomials of the same order. In particular, our transformation maps quadratic polynomials into quadratic pseudo-Boolean polynomials.

 The procedures above make QA an alternative approach addressing all problems admitting formulation as ILP or IQP. 

%%%%%%%%%%%%%%%%%%%%%%%%%%%%%%%%%%%%%%%%%%%%%%%%%%%%%%%%%%%%%%%%%%%%%%%%%%%%%%%%
\subsection{TSP problem and its variants}\label{sec:tsp}
%%%%%%%%%%%%%%%%%%%%%%%%%%%%%%%%%%%%%%%%%%%%%%%%%%%%%%%%%%%%%%%%%%%%%%%%%%%%%%%%

Let $ G=(V,\vec E) $ be directed graph where $ V=\{0,1,\dots,n\} $ is the set of nodes and $ \vec E \subset V\times V $ is the set of arcs. 

A \textbf{tour} consists of a sequence of vertices and edges, where the edges connect the adjacent vertices in the sequence, and no edge is repeated. A tour that visits each node exactly once is called a \textbf{Hamiltonian cycle}. For every pair of nodes $ (u,v) $, one can associate the cost of travelling from node $ u $ to $ v $ which is denoted by $ c_{uv} $. Finding a Hamiltonian cycle that minimizes the total cost of traveling between the nodes is known as the \textbf{Travelling Salesman Problem} (TSP).

Among the many generalizations of TSP, we will focus on \textbf{Travelling Salesman Problem with Time Windows} (TSPTW). Consider a vehicle that starts from the depot labeled by 0, visits each city, and returns to the depot. For each city $ v $, there is an associated service time which is included in the cost of the arc outgoing from $ v $ and a time window $ [e_v,l_v] $ such that the city $ v $ should be visited within the time window, where $e_v$ is the earliest start time and $l_v$ is the due time for city $v$. If the vehicle arrives at city $ v $ before $ e_v $,  the vehicle should wait. TSPTW aims to find a Hamiltonian cycle that minimizes the total cost and satisfies the time window constraints. Another objective in this setup would be to minimize the total completion time of the tour, known as the \textbf{Makespan Problem with Time Windows} (MPTW).

Both exact and heuristic algorithms have been proposed for TSPTW and its variants. Some of the first approaches include solutions to the MPTW problem. In \cite{baker1983exact}, a branch and bound procedure is utilized, and a non-linear program is formulated. \cite{langevin1993two} presents an integer program using a commodity flow formulation for both problems. In 2012, Baldacci et al. proposed an algorithm that outperformed the existing exact solutions using dynamic programming  \cite{baldacci2012new}. Some other dynamic programming approaches include the works of \cite{christofides1981state,dumas1995optimal}. The references \cite{pesant1998exact,focacci2002hybrid} provide exact solutions to the problem using constraint programming. A more recent study combining constraint programming and reinforcement learning is presented in \cite{cappart2020combining}. 

The first and the only attempt in solving the TSPTW problem using quantum algorithms is by Papalitsas et al. \cite{papalitsas2019qubo}, which we will discuss in more detail in the following section. There has also been some ongoing research on using quantum algorithms to solve the TSP problem and we can mention the various attempts of QAOA \cite{hadfield2017quantum,hadfield2019quantum,glos2020space} and QA \cite{martovnak2004quantum,santoro2006optimization}. Some other related work includes \cite{borowski2020new, irie2019quantum}, which use QA to solve vehicle routing problem, a generalization of TSP to multiple vehicles.  

%%%%%%%%%%%%%%%%%%%%%%%%%%%%%%%%%%%%%%%%%%%%%%%%%%%%%%%%%%%%%%%%%%%%%%%%%%%%%%%%
\section{Unconstrained binary models for TSPTW} \label{sec:formulation}
%%%%%%%%%%%%%%%%%%%%%%%%%%%%%%%%%%%%%%%%%%%%%%%%%%%%%%%%%%%%%%%%%%%%%%%%%%%%%%%%

A Hamiltonian cycle is a feasible solution for the TSPTW problem if the vehicle obeys the time window constraints of each city while visiting the cities in the order imposed by the cycle. The optimal solution is the feasible solution with the least cost. Given a cycle, let's investigate the tour of the vehicle in more detail. The vehicle leaves the depot immediately and moves to the first city on the tour. Then, there are three possibilities: If the vehicle enters the city within its time windows (if the arrival time is between the earliest start time and due time), then service is done immediately, and the vehicle moves to the next city. If the vehicle arrives before the earliest start time of the city, then it waits until the earliest start time and then moves to the next city. Finally, if the vehicle arrives later than the due time of the city, then the cycle is infeasible. After this procedure is repeated for each city on the cycle, the vehicle returns to the depot. Note that given the cycle, one can calculate the waiting times and services times using a classical procedure.  Hence, we are specifically interested in the Hamiltonian cycle that is a feasible or optimal solution to the problem. We will refer to such cycles as feasible routes and optimal routes respectively.

To describe the tour of the vehicle, we will introduce the concepts of arrival and waiting time for city $ v $ which will be denoted by $\alpha_v$ and $\nu_v$ respectively. We can describe the arrival times using a recurrence relation satisfying the following conditions:
\begin{enumerate}
	\item \emph{Initialization constraint}: Arrival $\alpha_v$ to the first visited city $v$ equals $c_{0v}$.
	\item \emph{Recurrence constraint}: If $w$ and $v$ are consecutive cities in the tour, $\alpha_v = \alpha_w + \nu_w+ c_{wv}$.
	\item \emph{Service constraint}: $\alpha_v + \nu _v$ (Service time for city $v$) is between $e_v$ and $l_v$.
\end{enumerate}
The constraints given above will form the backbone of the formulations we will present in the following subsections. The first one is the corrected and generalized version of the QUBO formulation presented in \cite{papalitsas2019qubo}. The second one is a HOBO formulation that is based on the standard QUBO formulation for the TSP problem given in \cite{lucas2014ising}. The third representation is a QUBO model based on the ILP formulated in \cite{kara2013new}.

%%%%%%%%%%%%%%%%%%%%%%%%%%%%%%%%%%%%%%%%%%%%%%%%%%%%%%%%%%%%%%%%%%%%%%%%%%%%%%%%
\subsection{Edge-based formulation}
%%%%%%%%%%%%%%%%%%%%%%%%%%%%%%%%%%%%%%%%%%%%%%%%%%%%%%%%%%%%%%%%%%%%%%%%%%%%%%%%
 We start with edge-based formulation of the TSPTW inspired by~\cite{papalitsas2019qubo}. A QUBO formulation for the TSPTW problem is presented in \cite{papalitsas2019qubo} with a simplifying assumption of $ e_v=0 $ for all $ v \in V $. However, there is a flaw in the given formulation, as we will discuss next.

A tour is of the form $ p_0,p_1,\dots,p_{n+1} $, where $ p_0=p_{n+1}=0 $ so that the tour starts and ends at the depot. We will refer to $ p_i $ as the node at position $ i $ of the tour or the $ i $-th visited city. For each $ i=2, \dots, n $ and $ u,v \in V$ s.t. $u\neq v$, let the binary variables $ x_{u,v}^i $ be defined as
\begin{equation}
x_{u,v}^i =  \begin{cases}%
1,      & \text{nodes $ u, v $ are at consecutive positions $ i-1 $ and $ i $ in the Hamiltonian cycle,}\\
0, & \text{otherwise.}
\end{cases}
\end{equation}
Furthermore, we introduce variables $x_{0,v}^1$ ($x_{v,0}^{n+1}$) which equals 1
iff $v$ is the first (last) visited city. Since each variable $x_{u,v}^i$
indicates the occurrence of edge $(u,v)$ in the tour, we call it an edge-based model.

%%%%%%%%%%%%%%%%%%%%%%%%%%%%%%%%%%%%%%%%%%%%%%%%%%%%%%%%%%%%%%%%%%%%%%%%%%%%%%%%
\subsubsection{Route checking}
%%%%%%%%%%%%%%%%%%%%%%%%%%%%%%%%%%%%%%%%%%%%%%%%%%%%%%%%%%%%%%%%%%%%%%%%%%%%%%%%
To check whether the given bit assignment is a Hamiltonian cycle, the authors of  \cite{papalitsas2019qubo} propose the Hamiltonian
\begin{align}
& \biggl ( 1 - \sum_{v=1}^n x_{0,v}^1 \biggr )^2 +  \sum_{i=2}^n \biggl (1- \sum_{\substack{u,v=1\\u \neq v}}^n x_{u,v}^i \biggr)^2 +  \biggl ( 1 - \sum_{v=1}^n x_{v,0}^{n+1} \biggr )^2 \nonumber \\
 &\phantom{\ =}+  \sum_{u=1}^n \biggl (1- \biggl ( x_{u,0}^{n+1} + \sum_{i=2}^n \sum_{\substack{v=1\\v \neq u}}^n x_{u,v}^i \biggr ) \biggr ) ^2 \nonumber   \\
 &\phantom{\ =} + \sum_{v=1}^n \biggl (1- \biggl ( x_{0,v}^{1} + \sum_{i=2}^n \sum_{\substack{u=1\\u \neq v}}^n x_{u,v}^i \biggr ) \biggr ) ^2 . \label{eq:papalitsas}
\end{align}

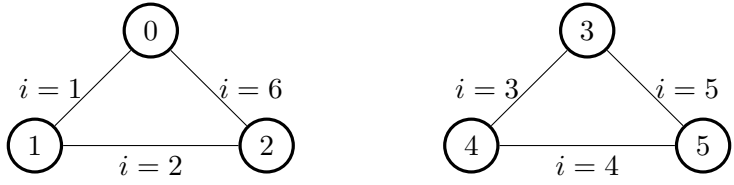
\begin{figure}[t!]
	\centering
	\begin{tikzpicture}[
	roundnode/.style={circle, draw=black, very thick, minimum size=7mm},
	squarednode/.style={rectangle, draw=red!60, fill=red!5, very thick, minimum size=5mm},
	]
	\node[roundnode] (node1) {0};
	\node[roundnode, below left= of node1] (node2) {1};
	\node[roundnode, below right= of node1] (node3) {2};
	
	\node[roundnode, right=5cm of node1] (node4) {3};
	\node[roundnode, below left= of node4] (node5) {4};
	\node[roundnode, below right= of node4] (node6) {5};
	
	\draw (node1) -- node[left] {$i=1$} (node2) --node[below] {$i=2$}(node3) --node[right] {$i=6$} (node1);
	
	\draw (node4) --node[left] {$i=3$} (node5) --node[below] {$i=4$} (node6) --node[right] {$i=5$} (node4);
	
	\end{tikzpicture}
	\caption{Undesired solution accepted in the formulation given in \cite{papalitsas2019qubo}.}
	\label{fig: undesired_solution}
\end{figure}

The first line imposes that for each $ i =1,\dots,n+1$, exactly one edge is traversed at each time step. The second line ensures that the vehicle leaves each node exactly once, and similarly, the third line checks whether each node is entered exactly once. This approach omits the subtour conditions which are required to ensure that the solution consists of a single closed tour. Taking into account the form of the Hamiltonian in Eq. \eqref{eq:papalitsas}, it is possible to find a solution in the form of two disjoint paths as presented in Figure \ref{fig: undesired_solution}.

To remove disjoint tours from the set of feasible solutions, we need to include an additional term in the Hamiltonian, ensuring that the city that is left and entered at consecutive times is the same. This leads to penalty term of the form 
\begin{equation}\label{eq:qubo-missing-cond-simple} 
\sum_{v=1}^n \biggl( 1 - \sum_{w=1}^n x_{0,v}^1 x_{v,w}^{2} \biggr)^2 
+ \sum_{i=2}^{n-1} \sum_{v=1}^n \biggl( 1 - \sum_{\substack{u,w=1\\u \neq w}}^n x_{u,v}^i x_{v,w}^{i+1} \biggr)^2 
+ \sum_{v=1}^n \biggl( 1 - \sum_{u=1}^n x_{u,v}^n x_{v,0}^{n+1} \biggr)^2.
\end{equation}

The above is not a QUBO anymore as we have terms of order 4. We claim that the squares can be removed, if exactly one city is visited at each step.

Let $i,v$ be an arbitrary pair of time step $ i $ and city $v$. Based on the condition from the first line of Eq.~\eqref{eq:papalitsas}, for given $i$ there exists exactly one $u'$ such that $x_{u',v}^i=1$. This transforms the formula inside the parenthesis into
\begin{equation}
1 - \sum_{\substack{u,w=1\\u \neq w}}^n x_{u,v}^i x_{v,w}^{i+1} = 1 - \sum_{\substack{w=1\\w \neq u'}}^n x_{u',v}^i x_{v,w}^{i+1} = 1 - \sum_{\substack{w=1\\w \neq u'}}^n  x_{v,w}^{i+1}.
\end{equation}
For the given $ v $, $x_{v,w}^{i+1}$ is either 0 for all $ w $ or following the reasoning above, there is exactly one $w'$ such that
$x_{v,w'}^{i+1}=1$, again based on Eq.~\eqref{eq:papalitsas}. Hence the expression inside the parenthesis is either equal to
0 or 1, and thus the square can be
omitted. Finally, we have the following condition:
\begin{equation}\label{qubo-correction}
\sum_{v=1}^n \biggl( 1 - \sum_{w=1}^n x_{0,v}^1 x_{v,w}^{2} \biggr) 
+ \sum_{i=2}^{n-1} \sum_{v=1}^n \biggl( 1 - \sum_{\substack{u,w=1\\u \neq w}}^n x_{u,v}^i x_{v,w}^{i+1} \biggr) 
+ \sum_{v=1}^n \biggl( 1 - \sum_{u=1}^n x_{u,v}^n x_{v,0}^{n+1} \biggr).
\end{equation}

Let us now show that the last line in Eq.~\eqref{eq:papalitsas} is not required anymore. Recall that the first line of Eq.~\eqref{eq:papalitsas} accounts for checking whether exactly one edge is traversed at each time step and the second line ensures that the vehicle leaves each node exactly once. Eq.~\eqref{qubo-correction} ensures that if $x_{v,w}^{i}=1$, then for some unique $u$ we have $x_{u,v}^{i-1}=1$. Together with the first and the second lines, this already enforces the condition that if the vehicle leaves node $ v $, it should have entered $ v $ in the previous time step, eliminating the necessity for having the third line in Eq.~\eqref{eq:papalitsas}.

In summary, we can check whether the resulting tour is a Hamiltonian cycle using the following Hamiltonian
\begin{align}\label{eq:hr}
H_{\mathcal R}&=
\biggl ( 1 - \sum_{v=1}^n x_{0,v}^1 \biggr )^2 +  \sum_{i=2}^n \biggl (1- \sum_{\substack{u,v=1\\u \neq v}}^n x_{u,v}^i \biggr)^2 +  \biggl ( 1 - \sum_{v=1}^n x_{v,0}^{n+1} \biggr )^2 \nonumber \\
&\phantom{\ =}+  \sum_{u=1}^n \biggl (1- \biggl ( x_{u,0}^{n+1} + \sum_{i=2}^n \sum_{\substack{v=1\\v \neq u}}^n x_{u,v}^i \biggr ) \biggr ) ^2 \nonumber   \\
&\phantom{\ =} + \sum_{v=1}^n \biggl( 1 - \sum_{w=1}^n x_{0,v}^1 x_{v,w}^{2} \biggr) 
+ \sum_{i=2}^{n-1} \sum_{v=1}^n \biggl( 1 - \sum_{\substack{u,w=1\\u \neq w}}^n x_{u,v}^i x_{v,w}^{i+1} \biggr) 
+ \sum_{v=1}^n \biggl( 1 - \sum_{u=1}^n x_{u,v}^n x_{v,0}^{n+1} \biggr).
\end{align}

%%%%%%%%%%%%%%%%%%%%%%%%%%%%%%%%%%%%%%%%%%%%%%%%%%%%%%%%%%%%%%%%%%%%%%%%%%%%%%%%
\subsubsection{Time-windows constraints}
%%%%%%%%%%%%%%%%%%%%%%%%%%%%%%%%%%%%%%%%%%%%%%%%%%%%%%%%%%%%%%%%%%%%%%%%%%%%%%%%

Recall that in the TSPTW problem, each city $ v $ has a time window $ [e_v,l_v]
$ where both $ e_v $ and $ l_v $ are integer variables. In the state-of-the-art
formulation of TSPTW \cite{papalitsas2019qubo}, it is assumed that $ e_v=0 $ for
all $ v $.  Next, we will improve the Hamiltonian so that arbitrary $0\leq e_v\leq l_v$ will be allowed.
Furthermore, instead of using one-hot encoding to express slack variables, we
will use binary encoding as we discussed in Section ~\ref{sec:qubo}, which will exponentially reduce the number of qubits representing the slack variables.

We start by defining the recurrence relation for arrival times, ensuring all three time windows constraints mentioned at the beginning of this section. For each $i=1,\dots, n+1$, let $A_i$ denote the arrival time to $i$-th visited
city ($A_{n+1}$ is the arrival time to the depot), and let $\omega_i$ be integer variables denoting the waiting time at $i$-th visited city. $A_i$ can be expressed using the following recurrence relation
\begin{equation}
A_1=\sum_{v=1}^n x_{0,v}^1 c_{0v}, \qquad
A_{i} = A_{i-1} + \omega_{i-1} + \sum_{\substack{u,v=1\\ u\neq v}}^n c_{uv} x_{u,v}^{i}, \label{eq:tw-ai-definition}
\end{equation}
 which can be expressed explicitly as
 \begin{equation}\label{eq: tw}
A_1=\sum_{v=1}^n x_{0,v}^1 c_{0v}, \qquad A_i = \sum_{t=1}^{i-1} \omega_t + \sum_{v=1}^n x_{0,v}^1 c_{0v} + \sum_{t=2}^{i} \sum_{\substack{u,v=1\\ u\neq v}}^n c_{uv} x_{u,v}^t  
 \end{equation}
for $i=2,\dots,n+1$. Note that the very definition of $A_i$ already implies initialization and recurrence constraints.

The service constraints take the form 
\begin{gather}
\sum_{v=1}^n x_{0,v}^1 e_{v} \leq A_1 +\omega_1 \leq \sum_{v=1}^n x_{0,v}^1 l_{v}, \label{eq:latestineq1} \\
 \sum_{\substack{u,v=1\\ u\neq v}}^n x_{u,v}^i e_v \leq  A_i +\omega_{i} \leq \sum_{\substack{u,v=1\\ u\neq v}}^n x_{u,v}^i l_v \label{eq:latestineq}
\end{gather}
for each $i=2,\dots,n$. Next we will prove that the inequalities in Eq.~\eqref{eq:latestineq1} and Eq.~\eqref{eq:latestineq} can be replaced by Eq.~\eqref{eq: tw1} and Eq.~\eqref{eq: tw2},
\begin{gather}
\sum_{v=1}^n x_{u,v}^i e_v \leq  A_1 +\omega_{1}, \quad  A_1 \leq \sum_{v=1}^n x_{0,v}^i l_v \label{eq: tw1}\\
\sum_{\substack{u,v=1\\ u\neq v}}^n x_{u,v}^i e_v \leq  A_i +\omega_{i}, \quad  A_i \leq \sum_{\substack{u,v=1\\ u\neq v}}^n x_{u,v}^i l_v \label{eq: tw2}
\end{gather}

If $ A_i $ and $ \omega_i $ satisfy the former, then they already satisfy the latter. For the converse, suppose that the latter is satisfied, yet $A_i + \omega_i$ is greater than the corresponding due time. Instead of $ \omega_i $, $ \omega_i^{'} $ can be chosen so that $  A_i+\omega_i^{'} $ is equal to the corresponding due time which is still greater than or equal to the corresponding earliest start time and the superfluous waiting time $ \omega_i^{'} -\omega_i $ can be moved to
$\omega_{i+1}$. Note that the number of bit assignments encoding the feasible routes increases, as we allow bit assignments encoding feasible routes to exist in the search space with different assignments to $ \omega $ without penalty.

To state the final form of the Hamiltonian, the inequality constraints will be included in the objective function and the integer variables $ \omega_i $ will be transformed into binary variables according to the procedure described in Sec.~\ref{sec:qubo}.

To convert the inequalities into equalities, we will use slack variables $\xi_{e,i},\xi_{l,i}$ for $e$- and $l$-dependent inequalities for time-step $i$. Trivially $\xi_{e,i}$ and $\xi_{l,i}$ are lower bounded by 0 for the case when inequalities are tightly satisfied. Let's first define a lower bound for the arrival times $ A_i $, which will be useful for the rest of the discussion 
\begin{equation}
\underline{A}_i = \min_{v=1,\dots,n}c_{0v} + \sum_{t=1}^{i-1}c^{(t)},
\end{equation}
and $c^{(1)},c^{(2)},\ldots$ is a non-decreasing sequence of all costs $c_{uv}$ for $ u\neq 0 $.\footnote{One could even improve the bound above into shortest path from depot to any vertex of distance $k$.} 

For the $ l $-dependent inequalities, we define the the upper bound $ \bar \xi_{l,i} $ for the slack variables $\xi_{l,i}$ as
\begin{equation}
\bar \xi_{l,i} \coloneqq -(\underline{A}_i - \max_{v=1,\dots,n} l_v)= \max_{v=1,\dots,n} l_v - \min_{v=1,\dots,n}c_{0v} - \sum_{t=1}^{i-1}c^{(t)}  . 
\end{equation}

For the $e$-dependent inequalities, the slack variables should compensate the inequality when the arrival time $ A_i $ is greater than the earliest start time $ e_v $ of the corresponding city. Otherwise, the vehicle has to wait until the earliest start time and $ A_i+\omega_i=e_v $ which results in an equality. Since the arrival time should be less than or equal to the due time of the corresponding city, we have the following upper bound for the slack variables independent of the time point $ i $
\begin{equation}
\bar \xi_{e,i} \equiv \bar \xi_{e} \coloneqq  \max_{v=1,\dots,n}(l_v-e_v)
\end{equation} 

Next, we replace the integer variables by binary variables. Since the upper bounds for the slack variables are already discussed, let us focus on the waiting times. For each $ i $,  $ 0 \leq \omega_i $ and we have the equality when vehicle arrives later than the earliest start time and gives the service immediately. For the upper bound, note that if the vehicle arrives early, it is sufficient for the vehicle to wait until the earliest start time. Hence, for $ \omega_i $, we define the upper bound $\bar  \omega_i $ as
\begin{equation}
\bar \omega_i \coloneqq \max_{v=1,\dots,n} e_v - \underline{A}_i  = \max_{v=1,\dots,n} e_v - \min_{v=1,\dots,n}c_{0v} - \sum_{t=1}^{i-1}c^{(t)}.
\end{equation}

Now, we are ready to present the final form of the penalty Hamiltonian
\begin{equation}
\begin{split}
H_{\mathcal TW} & \coloneqq  \biggl(\sum_{v=1}^n x_{0,v}^1 e_{v} - A_1 -E_0^{\bar \omega_i}(\omega_1) + E^{\bar \xi_e}_0(\xi_{e,1})\biggr)^2 + 
\biggl(A_1 - \sum_{v=1}^n x_{0,v}^1 l_{v} + E^{\bar \xi_{l,1}}_0(\xi_{l,1})\biggr)^2\\
&\phantom{\ \coloneqq } + \sum_{i=2}^{n}\biggl(\sum_{\substack{u,v=1\\ u\neq v}}^n x_{u,v}^i e_v -  A_i -E_0^{\bar \omega_i}(\omega_i) + E^{\bar \xi_e}_0(\xi_{e,i}) \biggr)^2 \\
&\phantom{\ \coloneqq } + \sum_{i=2}^{n}\biggl(A_i-\sum_{\substack{u,v=1\\ u\neq v}}^n x_{u,v}^i l_v + E^{\bar \xi_{l,i}}_0(\xi_{l,i})\biggr)^2 \label{eq:qubo-htw-final}.
\end{split}
\end{equation}

%%%%%%%%%%%%%%%%%%%%%%%%%%%%%%%%%%%%%%%%%%%%%%%%%%%%%%%%%%%%%%%%%%%%%%%%%%%%%%%%
\subsubsection{Objective Hamiltonian and representation cost}
%%%%%%%%%%%%%%%%%%%%%%%%%%%%%%%%%%%%%%%%%%%%%%%%%%%%%%%%%%%%%%%%%%%%%%%%%%%%%%%%
In this paper we focus on TSPTW problem, where the objective is to minimize the total cost of the tour. Hence, the objective Hamiltonian takes the form
\begin{equation}
H_{\mathcal C}^{\rm TSPTW} \coloneqq \sum_{v=1}^n c_{0v} x_{0,v}^1  + \sum_{i=2}^{n} \sum_{\substack{u,v=1\\ u\neq v}}^n c_{uv} x_{u,v}^i + \sum_{v=1}^n c_{u0} x_{u,0}^{n+1} .
\end{equation}
Hence, the QUBO Hamiltonian for the TSPTW  problem can be expressed as
\begin{equation}
H_{\text{TSPTW}} \coloneqq P_1 H_{\mathcal R}+ P_2 H_{\mathcal TW}+ P_3 H_{\mathcal C}^{TSPTW},
\end{equation}
where $H_{\mathcal R}$ and $H_{\mathcal TW}$ are defined as in Eqs.~\eqref{eq:hr} and~\eqref{eq:qubo-htw-final}  and $ P_1,P_2,P_3 $ are the constants which need to
be adjusted.

Let us calculate the number of qubits required for the formulation. For each $ i=2,\dots,n $ and $ u,v $ different than 0, there exist variables of the form $ x_{u,v}^i $, a total of $ n(n-1)^2 $ variables. For the case $ i=1 $ and $ i=n+1 $, there are $ 2n $ additional variables of the form $ x_{0,v}^1 $ and $ x_{v,0}^{n+1} $. For the binary representation of the variables, $\omega_i$, $\xi_{e,i}$ and $\xi_{l,i}$, the number of required qubits are $\lfloor \log_2(\bar \omega)\rfloor$ +1, $\lfloor \log_2(\overline\xi_{e,i})\rfloor+1$ and $\lfloor \log_2(\overline\xi_{l,i})\rfloor +1$ respectively. Let's define $\delta \coloneqq\lfloor \log_2(\max_vl_v)\rfloor +1 $, noting that $ \max_vl_v $ is an upper bound for $ \omega_i $, $ \xi_{e,i} $, and $ \xi_{l,i} $. Hence, in total the number of required qubits is at most 
\begin{equation}
n^3 -2n^2 +3n +3n\delta = \order{n^3 + n\delta}.
\end{equation}
In general, $ \delta $ grows at least linearly by $ n $ since it is an upper bound for the arrival time to the $ n $-th city.

%%%%%%%%%%%%%%%%%%%%%%%%%%%%%%%%%%%%%%%%%%%%%%%%%%%%%%%%%%%%%%%%%%%%%%%%%%%%%%%%
\subsection{Node-based formulation} \label{sec:node-based}
%%%%%%%%%%%%%%%%%%%%%%%%%%%%%%%%%%%%%%%%%%%%%%%%%%%%%%%%%%%%%%%%%%%%%%%%%%%%%%%%

In this section, we will present a node-based formulation for the TSPTW problem. While edge-based formulation focuses on the variables of the form $x_{u,v}^i$, here we will define variables using the original idea presented for the TSP problem in \cite{lucas2014ising}. For each $ i=1, \dots ,n $ and $v=1,\dots,n$, let's define the binary variables $x_{v}^i $ such that
\begin{equation}
x_{v}^i =  \begin{cases}%
1,      & \text{node $ v $ is visited at $ i $-th step in the tour}\\
0, & \text{otherwise.}
\end{cases}
\end{equation}
Note that there is no need to define variables
for $i=0$  and $n+1$ as $ x_{v}^0= x_{v}^{n+1}=1 $. Since the binary variables now represent the node visited at each time step, the formulation is called node-based.

To check whether the tour is a Hamiltonian cycle, we can use the following Hamiltonian as defined in \cite{lucas2014ising}:
\begin{gather}
\tilde H_{\mathcal R}  \coloneqq \sum_{v=1}^n \biggl (1 - \sum_{i=1 }^{n} x_{v}^i \biggr )^2 +  \sum_{i=1}^{n} \biggl (1 - \sum_{v=1}^n x_{v}^i \biggr)^2.
\end{gather}

We define the Hamiltonian for the time windows constraints $ \tilde{H}_{\mathcal{TW}}^{i} $ analogously to the time windows constraints $ H_\mathcal{TW} $. Since $x_{u,v}^i=1$ iff both $x_u^{i-1}$ and $x_v^{i}$ are equal to 1, it is enough to replace the variables of the form $x_{u,v}^i  $ by $ x_u^{i-1}x_v^{i} $ and we obtain the following Hamiltonian:

\begin{equation}
\begin{split}
\tilde H_{\mathcal TW} & \coloneqq  \biggl(\sum_{v=1}^n x_{v}^1 e_{v} - \tilde A_1 -E_0^{\bar \omega_1}(\omega_1) + E^{\bar \xi_e}_0(\xi_{e,1})\biggr)^2 + 
\biggl(\tilde A_1 - \sum_{v=1}^n x_{v}^1 l_{v} + E^{\bar \xi_{l,1}}_0(\xi_{l,1})\biggr)^2\\
&\phantom{\ \coloneqq } + \sum_{i=2}^{n}\biggl(\sum_{\substack{u,v=1\\ u\neq v}}^n x_{u}^{i-1} x_{v}^i e_v -  \tilde A_i -E_0^{\bar \omega_i}(\omega_i) + E^{\bar \xi_e}_0(\xi_{e,i}) \biggr)^2 \\
&\phantom{\ \coloneqq } + \sum_{i=2}^{n}\biggl(\tilde A_i-\sum_{\substack{u,v=1\\ u\neq v}}^n x_{u}^{i-1} x_{v}^i l_v + E^{\bar \xi_{l,i}}_0(\xi_{l,i})\biggr)^2, \label{eq:hobo-htw-final}
\end{split}
\end{equation}
where 
 \begin{equation}
\tilde A_1=\sum_{v=1}^n x_{v}^1 c_{0v}, \qquad \tilde  A_i = \sum_{t=1}^{i-1} \omega_t + \sum_{v=1}^n x_{v}^1 c_{0v} + \sum_{t=2}^{i} \sum_{\substack{u,v=1\\ u\neq v}}^n c_{uv} x_{u}^{t-1} x_{v}^t \label{}
\end{equation}
Note that bounds on the slack variables or $\omega$ haven't changed. 

Finally the cost Hamiltonian takes the form
\begin{gather}
{\tilde H}_{\mathcal{C}}^{TSPTW} \coloneqq  \sum_{v=1}^n  c_{0v} x_v^1  +  \sum_{ \substack{u,v=1 \\ u\neq v}}^n c_{uv} \sum_{i=2 }^{n} x_{u}^{i-1}x_{v}^{i} +  \sum_{v=1}^n  c_{v0} x_{v}^n.
\end{gather}
and the Hamiltonian for the TSPTW problem is expressed as
\begin{equation}
\tilde{H}_{\text{TSPTW}} \coloneqq P_1 \tilde{H}_\mathcal{R} + P_2 \tilde{H}_\mathcal{TW} +  P_3 \tilde{H}_\mathcal{C} ^{TSPTW}.
\end{equation}

Let us now estimate the number of qubits required. For each $ i=1,\dots,n $ and $ v $ different than 0, there exist variables of the form $ x_{v}^i $, a total of $ n^2 $ variables. The number of qubits required to express $\omega$, $\xi_{e,i}$ and $\xi_{l,i}$ is the same as in the edge-based encoding. Thus, the number of required qubits is at most $\order{n^2+n\delta}$. 
\subsection{ILP Approach}

Finally, we will discuss a new QUBO formulation for the TSPTW problem based on the ILP model presented in \cite{kara2013new}. It is assumed that the costs between the cities satisfy the triangle inequality i.e. $ c_{uv} \leq c_{uw} + c_{wv} $ for all $ u,v,w = 1,\dots,n $.

The variables $ x_{uv} $ for all $ u,v =0,1,\dots, n $ such that $u\neq v$ are defined as
\begin{equation}
x_{u,v} =  \begin{cases}%
1      & \text{edge $ (u,v) $ appears in the tour},\\
0 & \text{otherwise.}
\end{cases}
\end{equation}
Note that unlike the edge-based formulation discussed previously, the time-step in which the edge is visited is not specified.

Let's recall the notation for the arrival time and waiting time at city $ v $. We denote the arrival time and waiting time for each $ v=1,\dots,n+1 $ by $ \alpha_v $ and $ \nu_v $ respectively. Note that by $\alpha_{n+1} $, we denote the arrival time to the depot at the end of the tour. Let $ \sigma_v $ denote the exact service time for city $ v $ (including the waiting time). We present the ILP formulation from \cite{kara2013new} in its entirety.
\begin{align}
\textrm{minimize}   \hspace{0.25in}&\sum_{\substack{u,v=0 \\ u \neq v}}^n c_{uv}x_{u,v}  \\
\textrm{subject to} \hspace{0.2in} & \sum_{u=0}^n x_{u,v}=1~~ &v=0,1,\dots,n \\
& \sum_{v=0}^n x_{u,v}=1~~ &u=0,1,\dots,n \\
& \sigma_v \geq e_v~~ &v=1,\dots,n \label{eq:sigma-low}\\
& \sigma_v \leq l_v~~ &v=1,\dots,n \label{eq:sigma-high}\\
& \sigma_v + c_{v0}x_{v,0} \leq \sum_{\substack{u,v=0 \\ u \neq v}}^n c_{uv}x_{u,v} + \sum_{v=1}^n \nu_v~~ &v=1,\dots,n \label{eq:superfluous} \\
&\alpha_v - c_{0v}x_{0,v} \geq 0 ~~ &v=1,\dots,n \\
& \alpha_v + (l_v - c_{0v})x_{0,v} \leq l_v~~ &v=1,\dots,n \\
& \sigma_v = \alpha_v + \nu_v~~ &v=1,\dots,n \label{eq:service}\\
& \sigma_u - \alpha_v + (l_u - c_{0v} + c_{uv} ) x_{u,v} \leq l_u - c_{0v} ~~ &u\neq v;~ u,v=1,\dots,n \\
&\alpha_v - \sigma_u + (l_v - e_u- c_{uv} ) x_{u,v} \leq l_v - e_u	~~ &u\neq v;~ u,v=1,\dots,n\\[2ex]
&  x_{u,v}\in \{0,1\}  & u\neq v;~ u,v=1,\dots,n \\
&  \sigma_{v}\in \ZZ_{\geq 0},  \alpha_v \in \ZZ_{\geq 0},\nu_v \in \ZZ_{\geq 0}.    & v=1,\dots,n 
\end{align}

In ILP, the precedence is given to reducing the space of feasible solutions over the number of variables used and some constraints are added specifically for that purpose. Before transforming ILP problem into QUBO formulation, we can remove those constraints such as Eq.~\eqref{eq:superfluous}, as converting them to equalities would require additional slack variables. We will remove Eqs.~\eqref{eq:sigma-low} and~\eqref{eq:sigma-high} as they define the upper and lower bounds for the variables $ \sigma_v $ and those bounds will be utilized while converting the integer variables into binary. The variables $ \alpha_v $ will be replaced by $ \sigma_v - \nu_v $ and the Eq.~\eqref{eq:service} will be removed. We prefer to remove $ \alpha_v $ instead of $ \sigma_v $ or $ \nu_v $ as the range for the variables $ \alpha_v $ is larger and they are more qubit consuming. We express the simplified ILP problem as follows:
\begin{align}
\textrm{minimize} \hspace{0.25in} &\sum_{\substack{u,v=0 \\ u \neq v}}^n c_{uv}x_{u,v} \\
\textrm{subject to} \hspace{0.2in} & \sum_{u=0}^n x_{u,v}=1~~ &v=0,1,\dots,n \label{eq:route1} \\
& \sum_{v=0}^n x_{u,v}=1~~ &u=0,1,\dots,n \label{eq:route2} \\
& c_{0v}x_{0,v} \leq \sigma_v - \nu_v ~~ &v=1,\dots,n \label{eq:init-geq}\\
& \sigma_v - \nu_v + (l_v - c_{0v})x_{0,v} \leq l_v~~ &v=1,\dots,n \label{eq:init-leq} \\
& \sigma_u - \sigma_v + \nu_v + (l_u - c_{0v} + c_{uv} ) x_{u,v} \leq l_u - c_{0v} ~~ &u\neq v;~ u,v=1,\dots,n \label{eq:orig27}\\
&\sigma_v - \nu_v - \sigma_u + (l_v - e_u- c_{uv} ) x_{u,v} \leq l_v - e_u	~~ &u\neq v;~ u,v=1,\dots,n \label{eq:orig28}\\[2ex]
&  x_{u,v}\in \{0,1\}  & u\neq v;~ u,v=1,\dots,n \\
&  \sigma_{v}\in \ZZ_{\geq 0}, e_v \leq \sigma_{v}\leq l_v & v=1,\dots,n \\
& \nu_v \in \ZZ_{\geq 0}  & v=1,\dots,n 
\end{align}

The equations  \eqref{eq:init-geq} and \eqref{eq:orig28} initialize the arrival time to the first visited city and also ensure that $ 0 \leq \alpha_v \leq l_v $ for the remaining cities. Note that due to triangle inequality, $ c_{0v} \leq \alpha_v \leq l_v $ is also correct.

Next, we will transform the inequalities into equalities using slack variables and convert those variables to binary. The upper bounds for the slack variables for the equations \eqref{eq:init-geq}-\eqref{eq:init-leq} can be defined as
\begin{equation}
\overline \xi_{\ref{eq:init-geq}, v} \coloneqq l_v \text{ and } \overline \xi_{\ref{eq:init-leq}, v} \coloneqq l_v - c_{0v} 
\end{equation}
noting that
\begin{align}
\xi_{\ref{eq:init-geq}, v} &\leq   - (\min\{c_{0v},0\}-  \sigma_v + \nu_v )=  \alpha_v \leq l_v,  \\
\xi_{\ref{eq:init-leq}, v} &\leq  -(\sigma_v -\nu_v  + \min\{l_v - c_{0v},0\} -l_v) =   l_v-\alpha_{v} \leq l_v - c_{0v}. 
\end{align}

The equations \eqref{eq:orig27} and \eqref{eq:orig28} are included in the model to ensure that $ \sigma_u + c_{u,v} = \alpha_v $ when $ x_{u,v}=1 $. If $ e_u + c_{u,v} \geq l_v $ for some pair of cities $ (u,v) $, then Equations \eqref{eq:orig27} and \eqref{eq:orig28} need to be discarded from the model. Hence, we will assume that  $ e_u + c_{u,v} \leq l_v $ when calculating the bounds for the slack variables. In Eq.~\eqref{eq:orig27} and Eq.~\eqref{eq:orig28}, the minimum is achieved when $ x_{u,v} $ is set to 0 since $ l_u - c_{0v} + c_{uv}  \geq 0 $ by triangle inequality and $ l_v-e_u-c_{uv} \geq 0 $ by our assumption. We can bound the variables $ \xi_{\ref{eq:orig27}, u,v} $ and $ \xi_{\ref{eq:orig28}, u,v} $ as
\begin{align}
\xi_{\ref{eq:orig27}, u,v} &\leq -(\sigma_u - \sigma_v +  \nu_v + \min\{l_u - c_{0v} + c_{uv} ,0\} - l_u + c_{0v}) \\
&= -\sigma_u + \alpha_v + l_u - c_{0v} \\
&\leq -e_u + l_v + l_u - c_{0v} \\
\xi_{\ref{eq:orig28}, u,v}& \leq -(\sigma_v - \nu_v - \sigma_u + \min\{l_v - e_u- c_{uv},0\} - l_v + e_u) \\
&= -\alpha_v + \sigma_u + l_v - e_u  ) \\
&\leq -c_{0v} + \l_u + l_v - e_u.
\end{align}
Hence, the upper bound for the slack variables are defined as
\begin{equation}
\overline \xi_{\ref{eq:orig27}, u,v} \coloneqq -e_u + l_v + l_u - c_{0v} \text{ and } \overline \xi_{\ref{eq:orig28}, u,v} \coloneqq -c_{0v} + \l_u + l_v - e_u.
\end{equation}

The integer variables $ \nu_v $ can be bounded such that $ 0 \leq \nu_v \leq \bar \nu_v $ implying
\begin{equation}
\bar \nu_v \coloneqq e_v - c_{0,v}.
\end{equation}

For completeness, let us write the full Hamiltonian based on the ILP above. We will apply penalty $P_1$ for the the constraints defined in Eq.~\eqref{eq:route1} and Eq~\eqref{eq:route2} and $ P_2 $ for the remaining constraints. For simplicity, we are including all the constraints even if $e_u+c_{u,v}>l_v$. The final Hamiltonian takes the form

\begin{equation}
\hat{H}_{\text{TSPTW}} \coloneqq P_1 \hat{H}_\mathcal{R} + P_2 \hat{H}_\mathcal{TW} +  P_3 \hat{H}_\mathcal{C} ^{TSPTW}.
\end{equation}

where 
\begin{align}
 \hat{H}_\mathcal{R} &= \left(\sum_{u=0}^n \left(1-\sum_{v=0}^n x_{u,v}\right)^2 + \sum_{v=0}^n\left(1-\sum_{u=0}^n x_{u,v}\right)^2\right) ,\\
 \hat{H}_\mathcal{TW} &= \left(\sum_{v=1}^n\left (E_{e_v}^{l_v}(\sigma_v) - E_0^{\bar \nu_v}(\nu_v) - c_{0v}x_{0,v}  - E_0^{\overline \xi_{\ref{eq:init-geq}, v}}(\xi_{\ref{eq:init-geq}, v})\right)^2\right.\\
&+\sum_{v=1}^n\left (E_{e_v}^{l_v}(\sigma_v) - E_0^{\bar \nu_v}(\nu_v) + (l_v - c_{0v})x_{0,v} - l_v  + E_0^{\overline \xi_{\ref{eq:init-leq}, v}}(\xi_{\ref{eq:init-leq}, v})\right)^2\\
&+\sum_{\substack{u,v=1\\u\neq v}}^n\left (E_{e_u}^{l_u}(\sigma_u) - E_{e_v}^{l_v}(\sigma_v) + E_{0}^{\bar \nu_v}(\nu_v) + (l_u - c_{0v} + c_{uv} ) x_{u,v} - l_u + c_{0v} + E_0^{\overline \xi_{\ref{eq:orig27}, u,v}}(\xi_{\ref{eq:orig27}, u,v})   \right)^2\\
&+\left .\sum_{\substack{u,v=1\\u\neq v}}^n\left (E_{e_v}^{l_v}(\sigma_v) - E_0^{\bar \nu_v}(\nu_v) - E_{e_u}^{l_u}(\sigma_u) + (l_v - e_u- c_{uv} ) x_{u,v} - l_v + e_u + E_0^{\overline \xi_{\ref{eq:orig28}, v}}(\xi_{\ref{eq:orig28}, u,v}) \right)^2 \right ),\\
\hat{H}_\mathcal{C} ^{TSPTW} &= \sum_{\substack{u,v=0 \\ u \neq v}} c_{uv}x_{u,v}.
\end{align}

Let us note that $  \hat{H}_\mathcal{R} $ does not account for the route constraints on its own as the subtour elimination constraints are included within the time windows constraints. Nevertheless, we use the notation $ \hat{H}_\mathcal{R} $ for consistency with the other formulations.

Let us calculate the number of qubits required for the given representation. $(n+1)^2$ qubits represent the variables $x_{u,v}$. For each $\nu_v$ and $\sigma_v$ for $v=1,\dots, n$, at most $\delta  $ qubits are required, a total of $2n\delta$ qubits. For each slack variable, at most $\order{\delta}$ qubits are required, and since we have $\order{n^2}$ inequalities, we will need $\order{n^2\delta}$ variables. Thus, in total we will need $\order{n^2 + n^2 \delta}$ qubits. 

\begin{comment}
Note that in general it is not possible to compare $\order{n| \edges | + n\delta}$, as it depends how $\delta$ is increasing with $n$ or $|\edges|$. However for sufficiently small $\delta$ (or low resolution time windows) we claim ILP will be less demanding in number of qubits. 
\end{comment}
%%%%%%%%%%%%%%%%%%%%%%%%%%%%%%%%%%%%%%%%%%%%%%%%%%%%%%%%%%%%%%%%%%%%%%%%%%%%%%%%
\subsection{Additional comments} \label{sec:comments}
%%%%%%%%%%%%%%%%%%%%%%%%%%%%%%%%%%%%%%%%%%%%%%%%%%%%%%%%%%%%%%%%%%%%%%%%%%%%%%%%
 All formulations work independently of whether the cost matrix is symmetric or not. It is assumed that $ e_v \geq c_{0v} $ for all $ \{v \in 1,\dots,n\} $ since the earliest start time for a city can not be smaller than $ c_{0v} $. In the ILP formulation, the variable $ \nu_v $ representing the waiting time for city $ v $ can be removed if $ e_v = c_{0v}  $. 

In the edge-based and ILP formulations, we can remove some variables when moving from one city to another is forbidden. This is the case when the graph is not complete or when $ e_u + c_{uv} \geq l_v$ for some $ (u,v)  \in \edges $. We can simply set $x_{u,v}^i=0$ or $ x_{u,v}=0 $ depending on the formulation and ignore those variables. In such a case, the number of required qubits is reduced to $\order{n|\edges| + n\delta}$ and $\order{|\edges| + |\edges|\delta}$ respectively for the edge-based and ILP formulations. Note that in the node-based formulation, we cannot remove any qubit even if traversing some arc is not possible. Furthermore, the corresponding inequalities given in Eqs.~\eqref{eq:orig27} and \eqref{eq:orig28} may be removed in the ILP formulation.

One can introduce alternative objective functions to encode different problems. For instance for the MPTW problem, we can use the following objective function for the edge-based formulation:
\begin{equation*}
H_{\mathcal C}^{\text{MPTW}} \coloneqq \sum_{i=1}^{n} \omega_i + H_{\mathcal C}^{\rm TSPTW}.
\end{equation*} 
Similar Hamiltonians can be defined for the node-based and ILP formulations as well. 

When the three models are compared, the number of required qubits is $\order{n^3+n\delta}$, $\order{n^2+n\delta}$ and $\order{n^2+n^2\delta}$ respectively for the edge-based, node-based and ILP formulations. When quadratization is performed to the HOBO formulation obtained through the node-based approach by replacing the product of variables $x_u^{t-1}x_v^{t} $ with $x_{uv}^t $, the resulting QUBO has asymptotically the same number of variables with that of the edge-based approach. Furthermore, the two formulations are similar in nature but the quadratized formulation involves additional constraints coming from the quadratization procedure itself. Further details on quadratization is given in the Appendix.  

To have an overview of the number of required variables for real problem instances, we calculated the number of required variables for the instances introduced in \cite{ascheuer1996hamiltonian} and the results are plotted in Figure \ref{fig:realinstancesvars}. Further details can be found in the Appendix. 

\begin{figure}[h]
	\centering
	\includegraphics[width=0.7\linewidth]{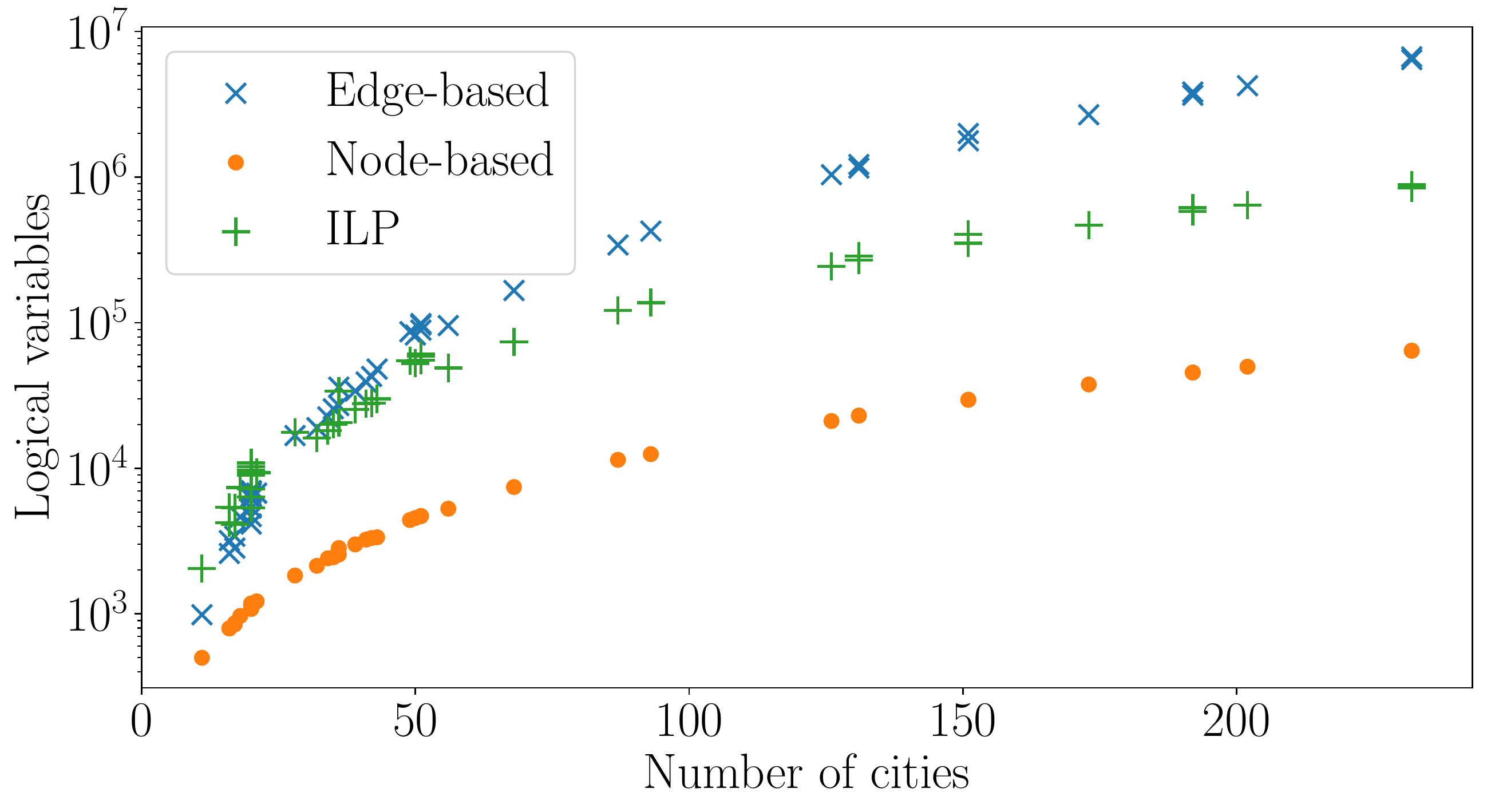}
	\caption{The number of variables required by different formulations for each instance from \cite{ascheuer1996hamiltonian}. }
	\label{fig:realinstancesvars}
\end{figure}

%Alternatively, if we have predefined another cost matrix denoting for example fuel cost spend by the vehicle, one should remove all $\nu_i$ in the formula above and replace each $c_{u,v}$ with the appropriate cost.

%%%%%%%%%%%%%%%%%%%%%%%%%%%%%%%%%%%%%%%%%%%%%%%%%%%%%%%%%%%%%%%%%%%%%%%%%%%%%%%%
\section{Results from the D-Wave machine} \label{sec:results}
%%%%%%%%%%%%%%%%%%%%%%%%%%%%%%%%%%%%%%%%%%%%%%%%%%%%%%%%%%%%%%%%%%%%%%%%%%%%%%%%

In this section, we will present the results conducted on the D-Wave Advantage QPU hardware. We ran several experiments using instances of small sizes to demonstrate the state-of-art capability of D-Wave systems. We used edge-based and ILP approaches to formulate the problems, as the HOBO formulation obtained using the node-based approach can be used only after performing quadratization as explained in Sec.~\ref{sec:comments}.  

Given that the penalty constants are well adjusted, the ground state encodes the optimal route with no penalties coming from the time-windows constraints and has the lowest energy. Note that, some bit assignments may encode an optimal (feasible) route yet violate some time windows constraints. Such assignments have larger energy than the bit assignment (if exist) with all time-constraints being satisfied. Throughout the section, we will use the term sample encoding optimal (feasible) route for the bit assignments encoding optimal (feasible) routes obtained by quantum or simulated annealing, regardless of whether they violate time-windows constraints or not. Such samples can be classified as optimal (feasible) since we are interested in the optimal (feasible) route and other variables can be calculated using a classical procedure mentioned in Section 3.

%%%%%%%%%%%%%%%%%%%%%%%%%%%%%%%%%%%%%%%%%%%%%%%%%%%%%%%%%%%%%%%%%%%%%%%%%%%%%%%%
\subsection{Instances}
%%%%%%%%%%%%%%%%%%%%%%%%%%%%%%%%%%%%%%%%%%%%%%%%%%%%%%%%%%%%%%%%%%%%%%%%%%%%%%%%

We created 10 random metric TSPTW instances for each $n \in \{3,4,5\}$ where $n$ is the number of cities. For the cost matrix, we picked random integers between 1-10, and for the earliest start times, we picked random integers between 1-20. For the latest start times, we picked random integers between the earliest start time of the city and 40. For the first 5 instances, the TSP and TSPTW solutions are the same, and for the remaining 5, they are different. An optimal route exists for each instance.   

%%%%%%%%%%%%%%%%%%%%%%%%%%%%%%%%%%%%%%%%%%%%%%%%%%%%%%%%%%%%%%%%%%%%%%%%%%%%%%%%
\subsection{Embedding}
%%%%%%%%%%%%%%%%%%%%%%%%%%%%%%%%%%%%%%%%%%%%%%%%%%%%%%%%%%%%%%%%%%%%%%%%%%%%%%%%
Before running a problem on D-Wave QPU, the QUBO formulation is converted into Ising formulation that consists of the linear and the quadratic terms. The quadratic terms represent the coupler strength between the qubits, and one can represent the variables by nodes and the quadratic terms by the edges. As D-Wave QPUs do not admit a fully connected topology, the variables can not be mapped directly to the physical qubits on the machine. Hence, each variable is represented by a set of qubits called the \textit{chain}, and the qubits in a chain are coupled strongly enough based on a parameter called \textit{chain strength} so that they end up in the same state. The existence of longer chains increases the error in the results. 

\begin{figure}[h]
\centering
\includegraphics[width=\linewidth]{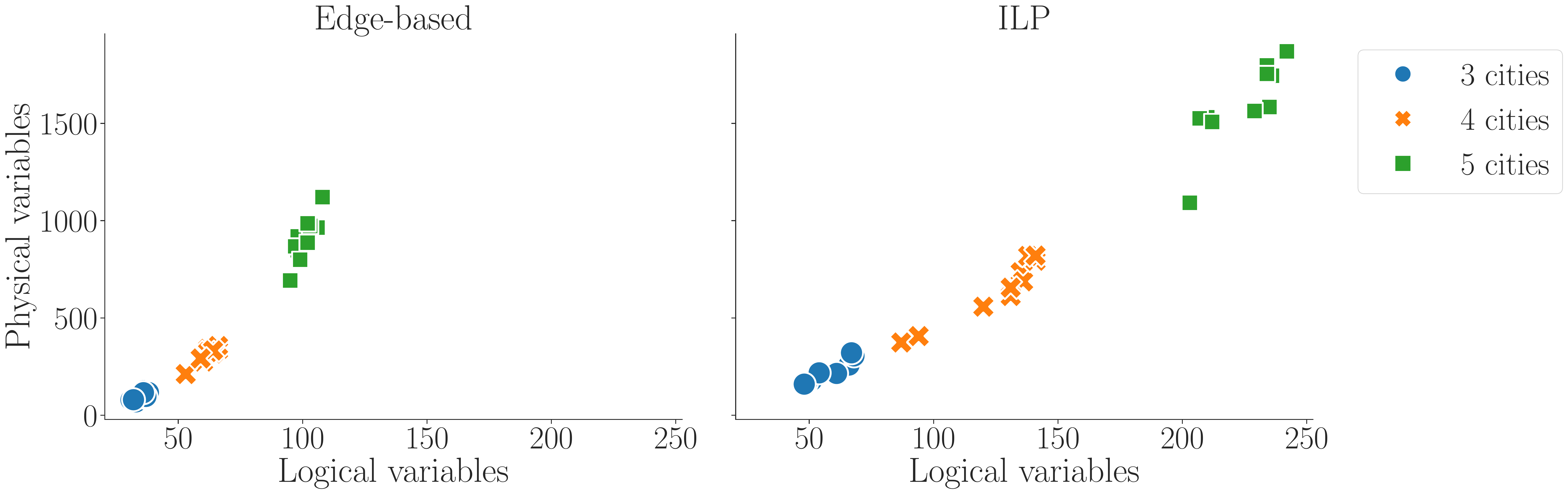} 
\caption{The number of logical and physical variables are plotted for each instance.}
\label{fig:variables}	
\end{figure}

This process of mapping the variables to the physical qubits is known as the \textit{minor-embedding} problem. We conducted our experiments on D-Wave Advantage that consists of 5640 qubits oriented in Pegasus graph topology \cite{boothby2020next}. For minor-embedding, we used the minorminer algorithm provided by D-Wave. For each instance, the number of logical and physical variables are given in Figure \ref{fig:variables} and the range for the maximum chain lengths for instances with different number of cities is given in Table \ref{tbl: chain} . 

\begin{table}[h]
	\centering
	\begin{tabular}{cll}
		\toprule
		Number of cities & Edge-based & ILP   \\ \midrule
		3                & 4--7        & 7--11  \\
		4                & 7--11       & 11--18 \\
		5                & 12--19      & 27--35 \\ \bottomrule
	\end{tabular}
\caption{Range for the maximum chain lengths for instances with different number of cities.}
\label{tbl: chain}	
\end{table}

%%%%%%%%%%%%%%%%%%%%%%%%%%%%%%%%%%%%%%%%%%%%%%%%%%%%%%%%%%%%%%%%%%%%%%%%%%%%%%%%
\subsection{Penalty constants}
%%%%%%%%%%%%%%%%%%%%%%%%%%%%%%%%%%%%%%%%%%%%%%%%%%%%%%%%%%%%%%%%%%%%%%%%%%%%%%%%
In the QUBO formulation, there exist objective Hamiltonian and Hamiltonians that account for the constraints. Associated with each Hamiltonian, there exist penalty constants whose magnitude depends on the penalty constant. When a constraint is violated, the Hamiltonian brings in a penalty to the energy, 

In our formulations, we have three constants, $ P_1 $, $ P_2 $ and $ P_3 $, corresponding to Hamiltonians that check whether the tour is a Hamiltonian cycle, that encodes the time windows constraints, and that accounts for the cost of the tour respectively. In the ILP formulation, subtour elimination constraints are included within the time windows constraints. In the original TSP problem, the only constants to be adjusted are $ P_1 $ and $ P_2 $ and it is enough to set $ 0 <  \mathcal{C}\cdot  P_2  <P_1$~\cite{lucas2014ising} where $  \mathcal{C} = \max_{v,w}c_{vw} $. The idea is to set $ P_1 $ large enough to ensure that the constraints are not violated in the favour of optimizing the cost. In the case of TSPTW, this requires a more detailed investigation as the penalty constant for the time windows needs to be adjusted as well, and multiple factors affect the penalty constant including the earliest and latest start times. If the penalty constants are not well adjusted, a bit assignment encoding an infeasible route may have lower energy than a bit assignment encoding the optimal route.  

To adjust the penalty constants, we set $ P_2=1 $ and parametrize $ P_1 $ and $ P_2 $ in terms of $ \mathcal{C} $. After the model is formulated, it is converted into Ising formulation, and both linear and quadratic terms are scaled to match the allowed range of the specific QPU. In order to determine the penalty constants, we used simulated annealing and tested a grid of penalty constants, and selected those which maximize the probability that the optimal route is observed in the sampleset for both formulations. For the simulated annealing experiments, we set the beta range (inverse temperature) as $ (5,100) $, the number of steps to 10000, and the number of samples as 100. Let us remark that using simulated annealing with the pre-knowledge of the optimal route is not applicable in practice, and there is not a one-to-one correspondence between the samples obtained by SA and QA experiments. Nevertheless, as the main aim of our experiments is demonstrating the capabilities and the limits of the D-Wave machines, this approach provides foresight into the selection of the penalty constants. 

\begin{figure}[ht]
	\centering
	
	\begin{subfigure}[b]{0.35\linewidth}
		\centering
		\includegraphics[width=1\linewidth]{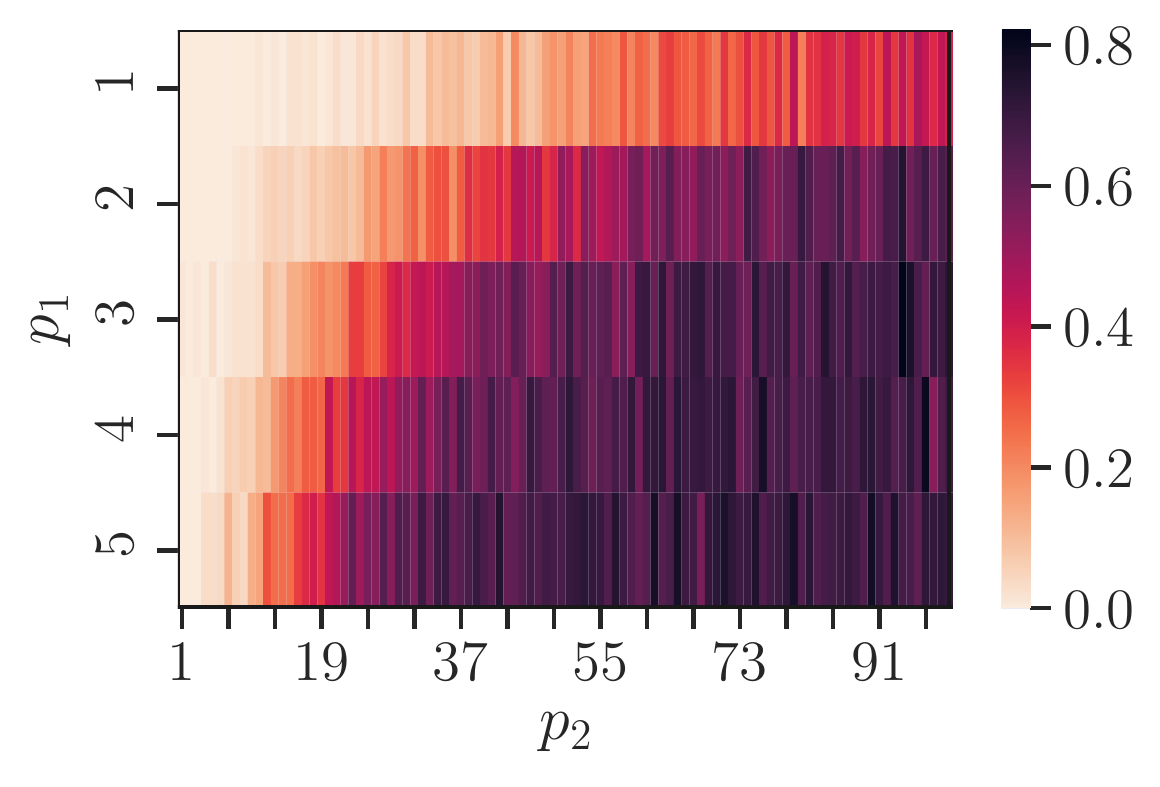} 
		\caption{}
	\end{subfigure}
	\begin{subfigure}[b]{0.35\linewidth}
	\centering
	\includegraphics[width=1\linewidth]{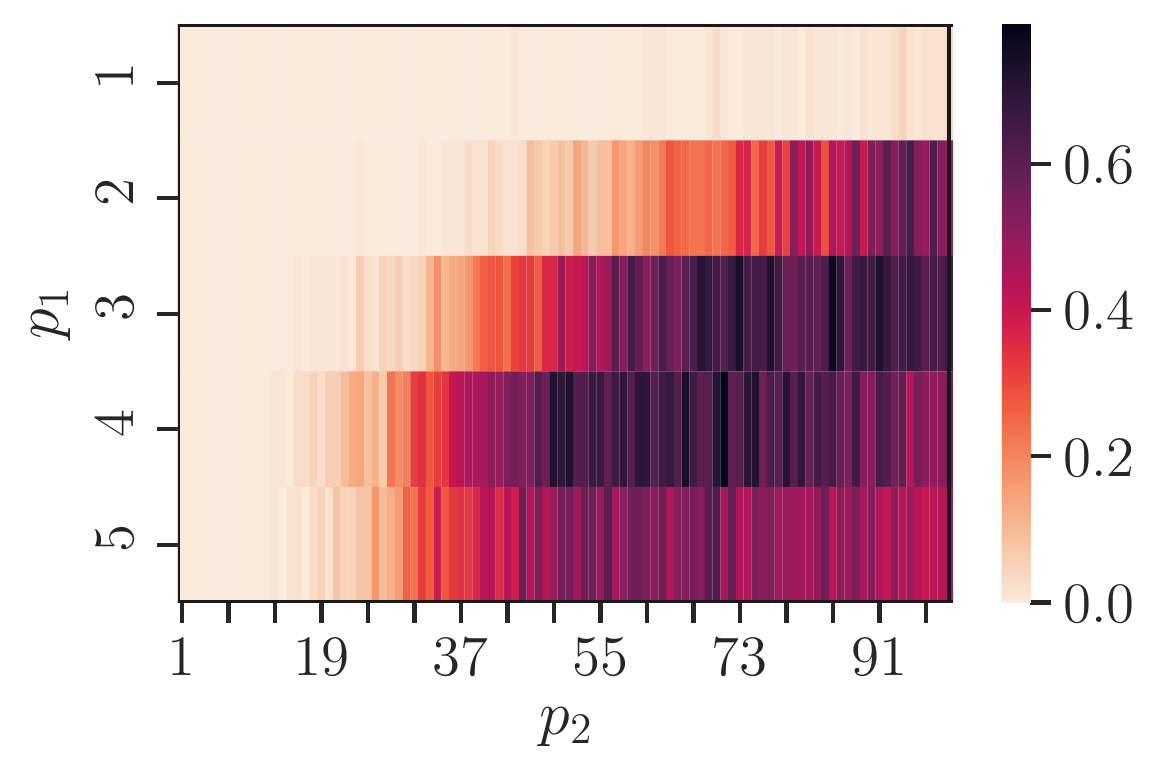} 
	\caption{}
\end{subfigure}

	\begin{subfigure}[b]{0.35\linewidth}
		\centering
		\includegraphics[width=1\linewidth]{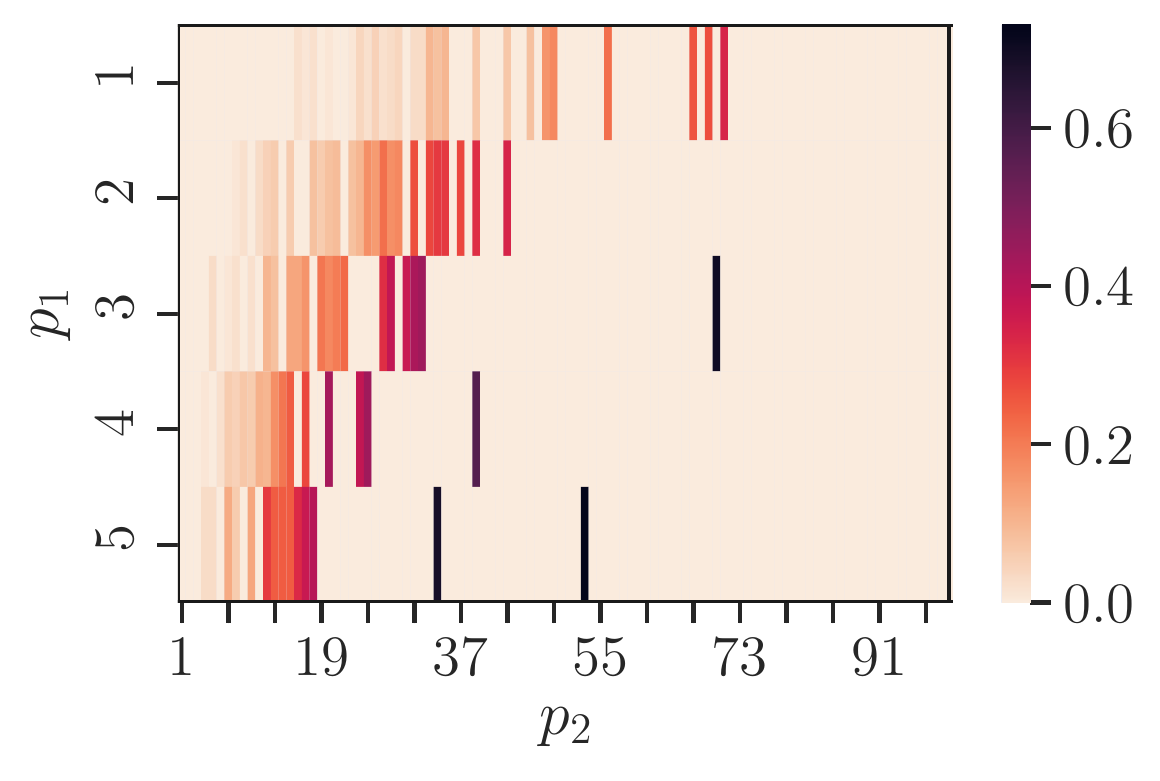} 
		\caption{}
	\end{subfigure}
	\begin{subfigure}[b]{0.35\linewidth}
	\centering
	\includegraphics[width=1\linewidth]{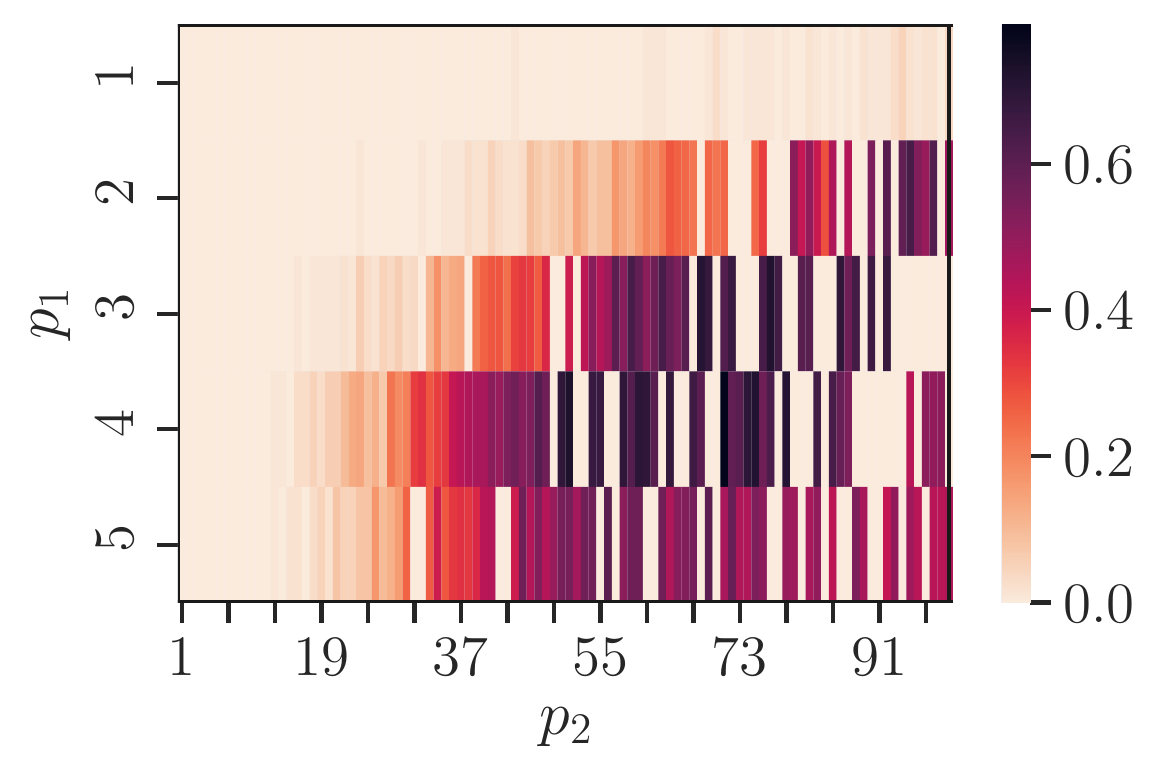} 
	\caption{}
\end{subfigure}
	\caption{The probability that a sample encoding the optimal route is observed for (a) edge-based and  (b) ILP fromulation for an instance with 4 cities. Simulated annealing is used for the probability estimation. In plots (c) and (d), the results are repeated except that we set probability to 0 if the least energy sample does not encode an optimal solution.}
	\label{fig:grid}
\end{figure}

In Figure \ref{fig:grid}, we plot the results of the SA experiments for different choices of $ p_1 $ and $ p_2 $ where $ P_1 = \mathcal{C} \cdot p_1 $ and $ P_2 = \mathcal{C}/{p_2}$. In the first two plots, the probability that a sample encoding the optimal route is observed within the sampleset is plotted for the edge-based and ILP formulations, respectively, for an instance with 4 cities. For some choice of penalty values, samples encoding non-optimal routes end up with lower energies than the samples encoding optimal route, yet the optimal route is also sampled with positive probability. In the last two plots, the probability for such penalty constant pairs is set to 0.

%%%%%%%%%%%%%%%%%%%%%%%%%%%%%%%%%%%%%%%%%%%%%%%%%%%%%%%%%%%%%%%%%%%%%%%%%%%%%%%%
\subsection{Optimization results}
%%%%%%%%%%%%%%%%%%%%%%%%%%%%%%%%%%%%%%%%%%%%%%%%%%%%%%%%%%%%%%%%%%%%%%%%%%%%%%%%
For each instance, we set $ P_1 $ and $ P_2$ based on the results of the SA experiments and run the experiments on D-Wave Advantage. For the experiments, we set the annealing time parameter as $ 50 \mu s $ and the number of samples as $ 1000$. We note that using shorter and longer annealing times did not provide any significant improvement. The chain strength is set between 1.4 and 2 using the following strategy: initially, the chain strength is set as 2 and it is decreased gradually if a satisfying outcome is not obtained, in the meantime keeping in mind that chain breaks should be avoided. 

\begin{figure}[ht]
		\centering
	\begin{subfigure}[b]{0.32\linewidth}
		\includegraphics[width=1\linewidth]{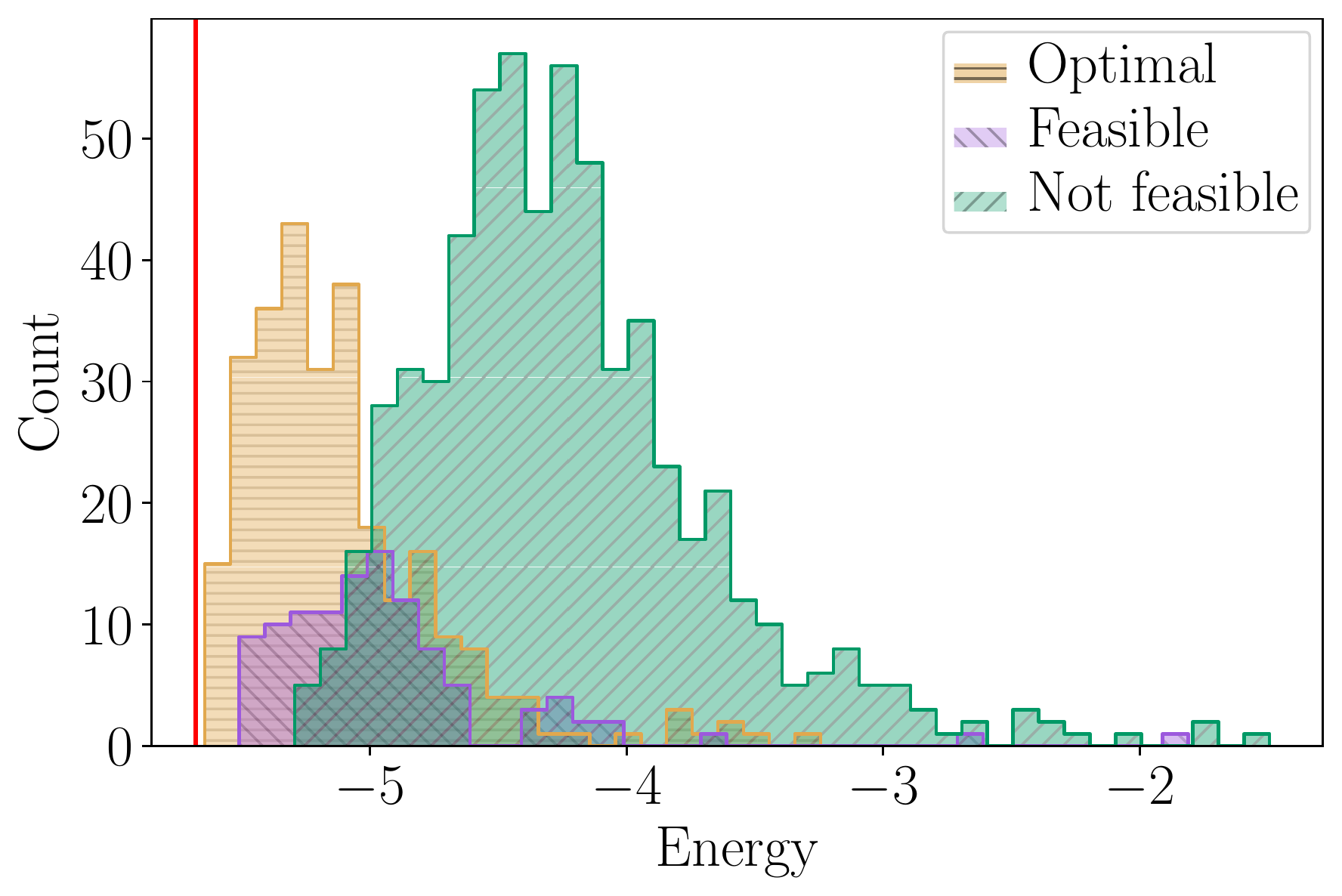} 
		\caption{An instance with 3 cities.}	
	\end{subfigure}\hspace{12pt}
	\begin{subfigure}[b]{0.32\linewidth}
		\centering
		\includegraphics[width=1\linewidth]{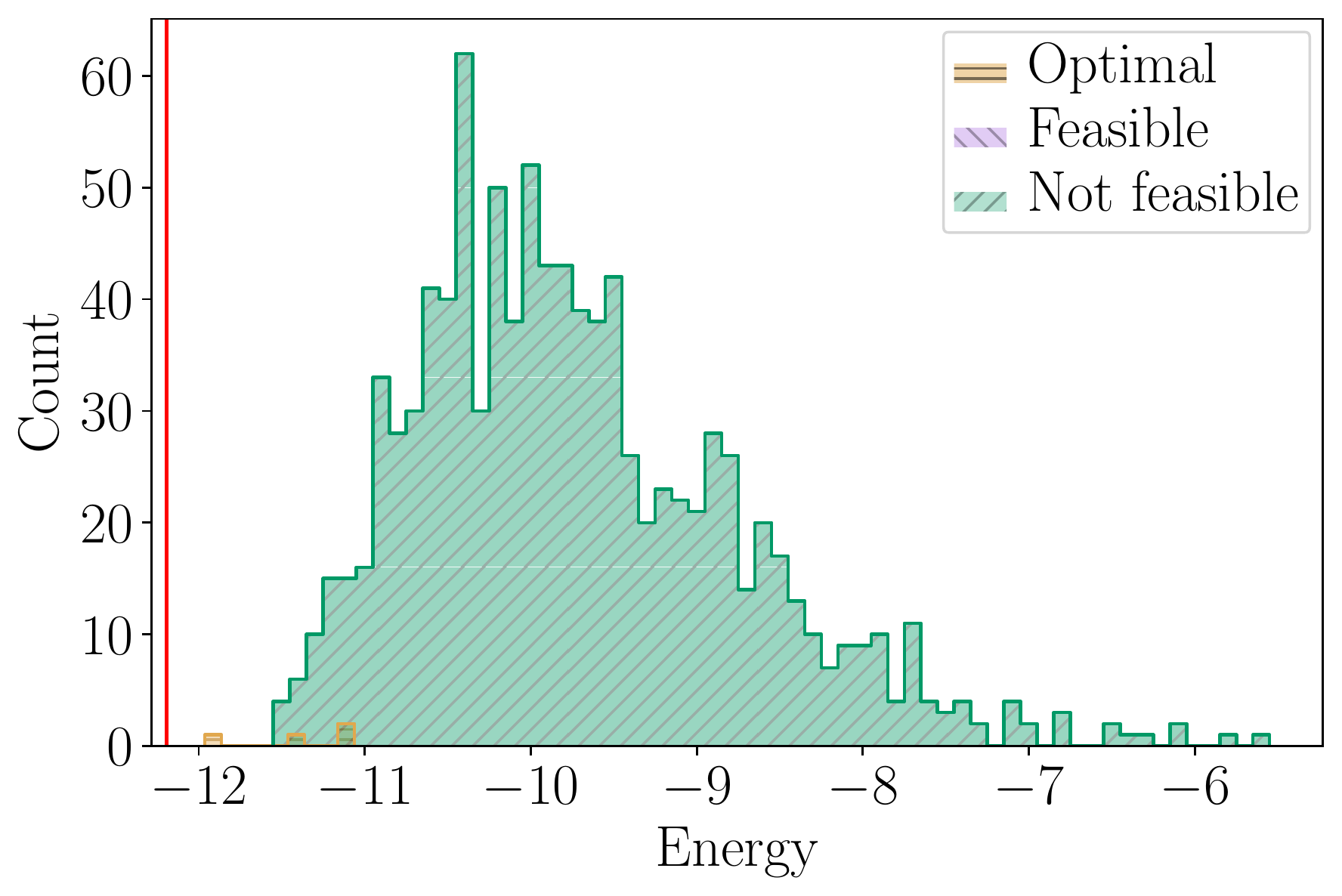} 
		\caption{An instance with 4 cities.}	
	\end{subfigure}

\caption{Histogram of the energies obtained from D-Wave using edge-based formulation.}	
\label{fig:hist}
\end{figure}

We begin with the edge-based formulation. Initially, we investigate whether the sample with the lowest energy encodes the optimal route or not for each instance. That was the case for all instances with 3 cities, whereas for instances with 4 cities the lowest energy sample did not encode the optimal route for three of the instances. In Figure \ref{fig:hist}, the histogram of the energies is plotted for an instance with 3 cities on the left, and for an instance with 4 cities on the right. The ground state energy is calculated and indicated by the red line on both plots. Note that even though the sample with the lowest energy encodes the optimal route on the right plot, its energy is higher than the ground state energy. This indicates that the bit assignment violates some time window constraints, nevertheless the correct route is found. 

\begin{figure}[ht]
		\centering
	\begin{subfigure}[b]{0.32\linewidth}
		\includegraphics[width=1\linewidth]{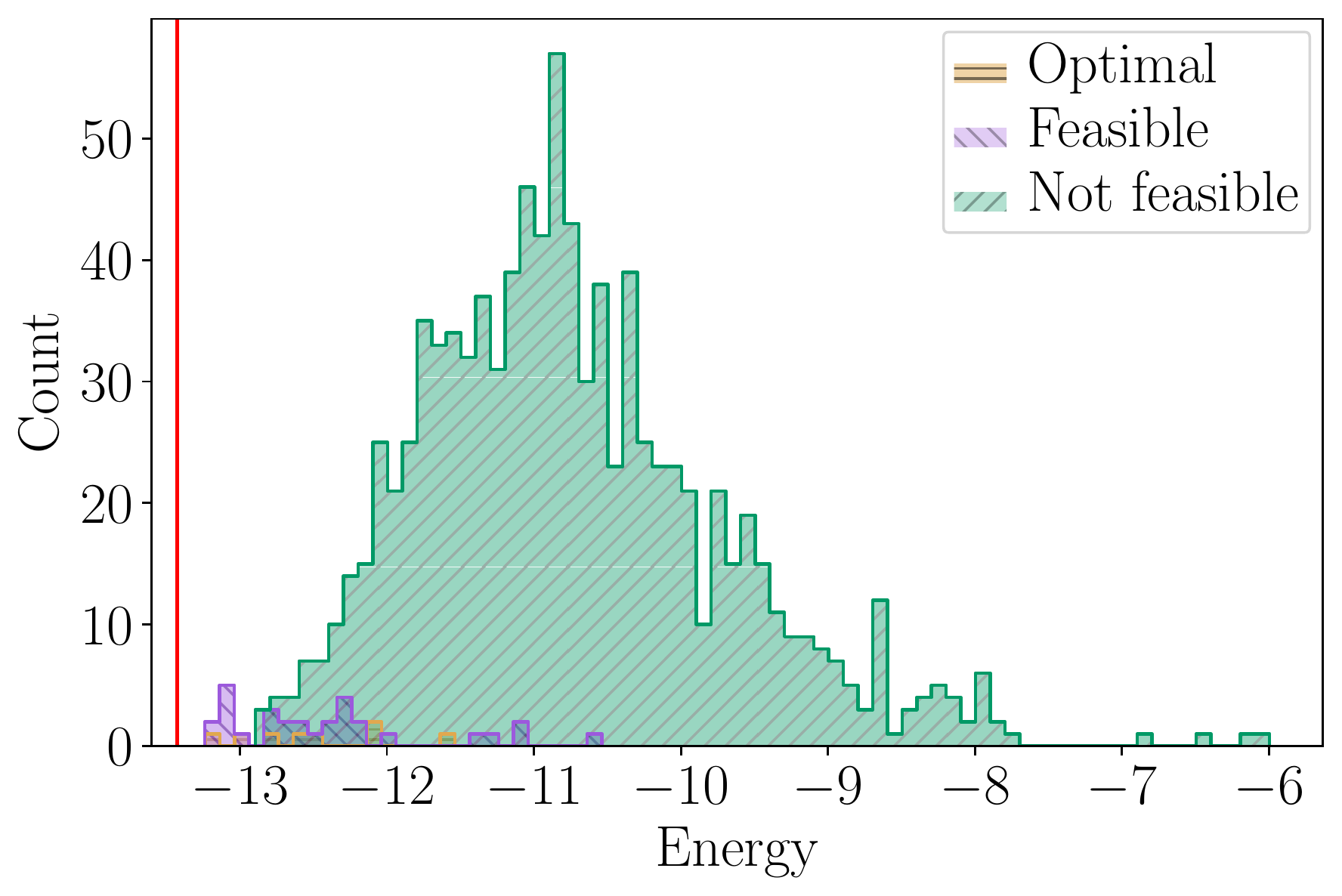} 
		\caption{An instance with 4 cities.}
	\end{subfigure}\hspace{12pt}
	\begin{subfigure}[b]{0.32\linewidth}	
		\centering
		\includegraphics[width=1\linewidth]{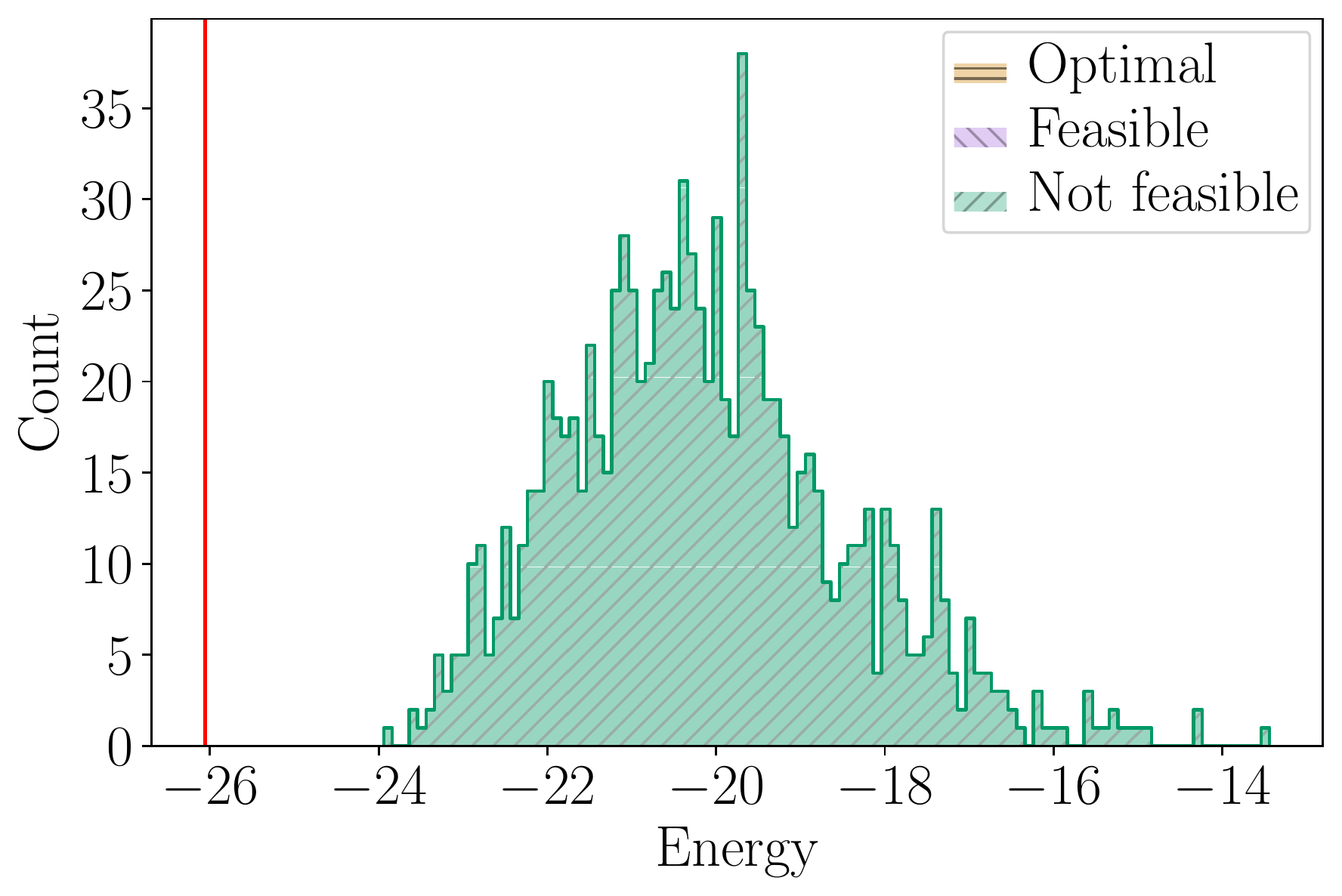} 
		\caption{An instance with 5 cities.}
	\end{subfigure}
\caption{Histogram of the energies obtained from D-Wave using edge-based formulation in which the lowest energy samples do not encode an optimal route. In plot (b) there are no feasible or optimal samples.}	
\label{fig:hist2}
\end{figure}

In Figure \ref{fig:hist2}, the histogram of the energies for two instances with 4 and 5 cities is plotted, in which the samples with the lowest energies do not encode the optimal route. For the instance with 4 cities, the sample with the lowest energy encodes a feasible route, and it is observed that the distribution of the energies is shifted to the right. This is even more apparent on the right plot for the instance with 5 cities as the gap between the sample with the lowest energy, and the ground state energy is larger. Indeed, this is true for all instances with 5 cities, and the existence of such a gap prevents sampling states which correspond to the optimal route.

\begin{figure}[ht]
	\begin{subfigure}[b]{0.32\linewidth}		
		\centering
		\includegraphics[width=1\linewidth]{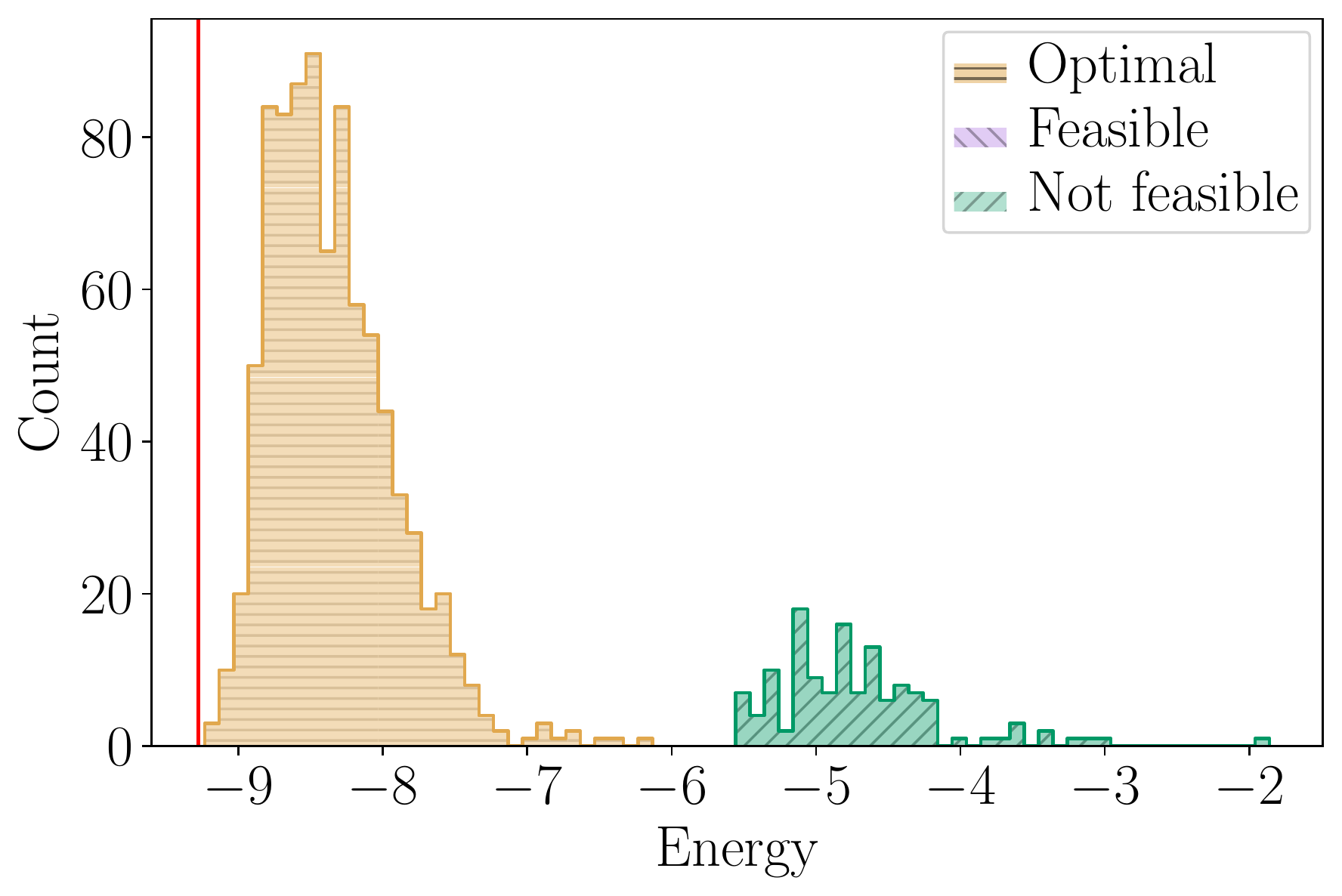}
		\caption{An instance with 3 cities.}
	\end{subfigure}
	\begin{subfigure}[b]{0.32\linewidth}
		\centering
		\includegraphics[width=1\linewidth]{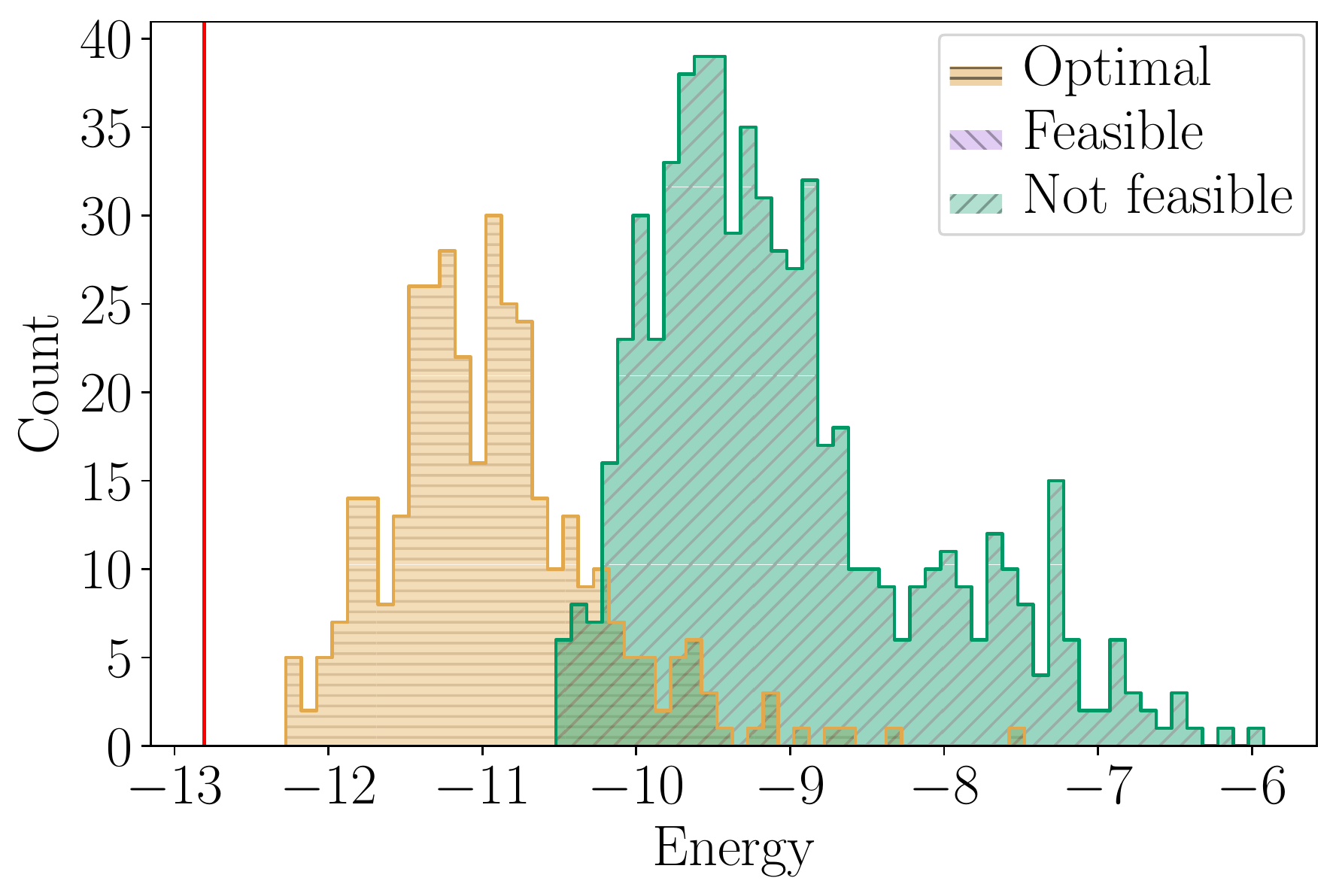}
		\caption{An instance with 4 cities.}
	\end{subfigure}
	\begin{subfigure}[b]{0.32\linewidth}
		\centering
		\includegraphics[width=1\linewidth]{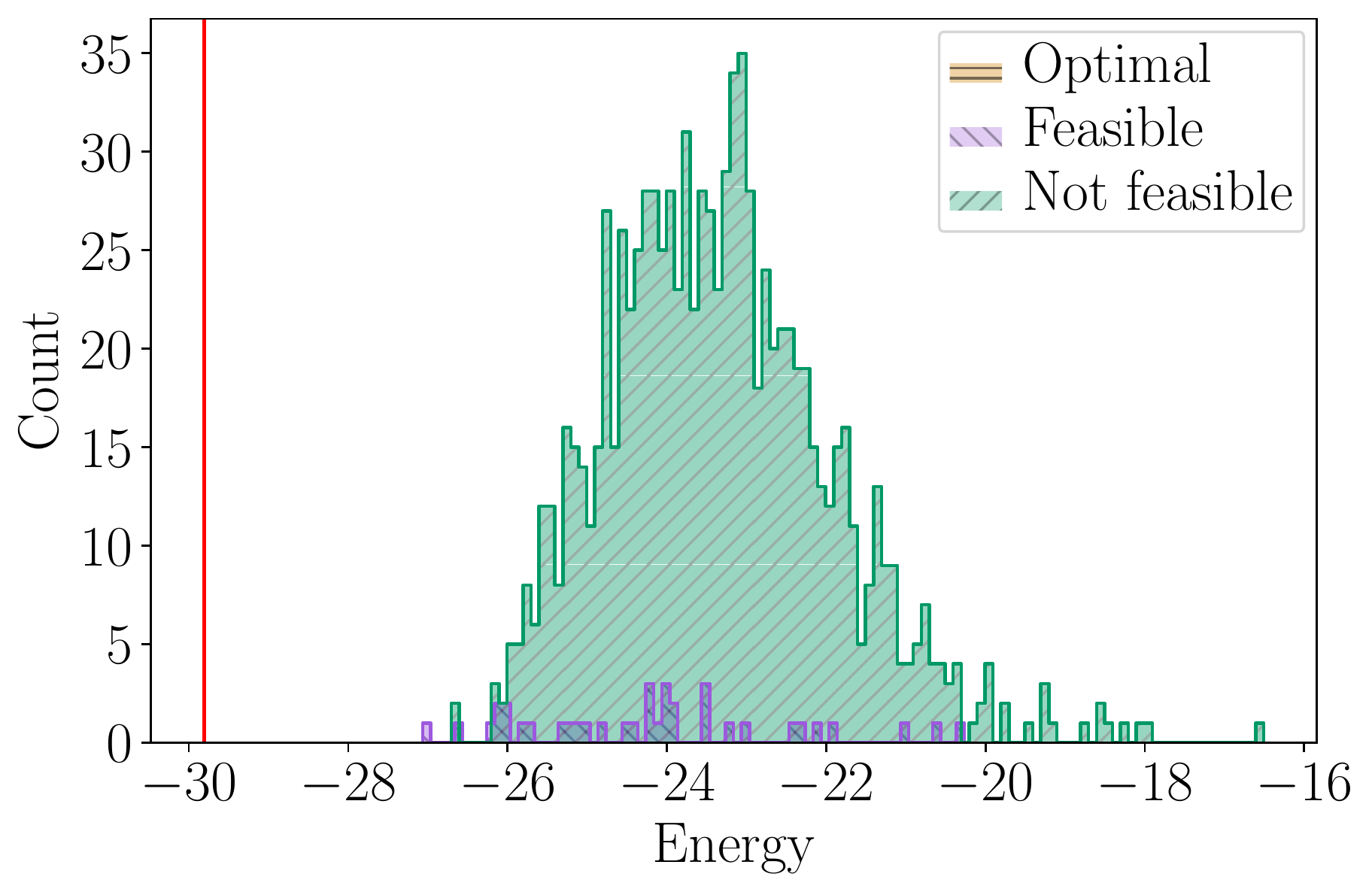}
		\caption{An instance with 5 cities.}
	\end{subfigure}
\caption{Histogram of the energies obtained from D-Wave using ILP formulation. In plots (a) and (b) there are no feasible samples and in plot (c), there are no optimal samples.}	
\label{fig:ilp}
\end{figure}

For the ILP formulation, the samples with the lowest energy encode optimal routes for all instances with 3 cities. For the instances with 4 cities, this was the case for only half of the instances. In Figure \ref{fig:ilp}, the first plot is the histogram for the energies obtained for an instance with 3 cities. Samples encoding optimal and feasible routes are separated clearly, and the distribution of the energies is not similar to that of the edge-based formulation. In the second plot, the histogram of the energies is plotted for an instance with 4 cities, and even though the sample with the lowest energy encodes the optimal route, its energy is larger than the ground state energy. In the third plot, it is seen that the distribution is even more shifted to the right for an instance with 5 cities.  

\begin{figure}[ht]
	\centering 

	\begin{subfigure}[b]{0.30\linewidth}
		\centering
		\includegraphics[height=3.5cm]{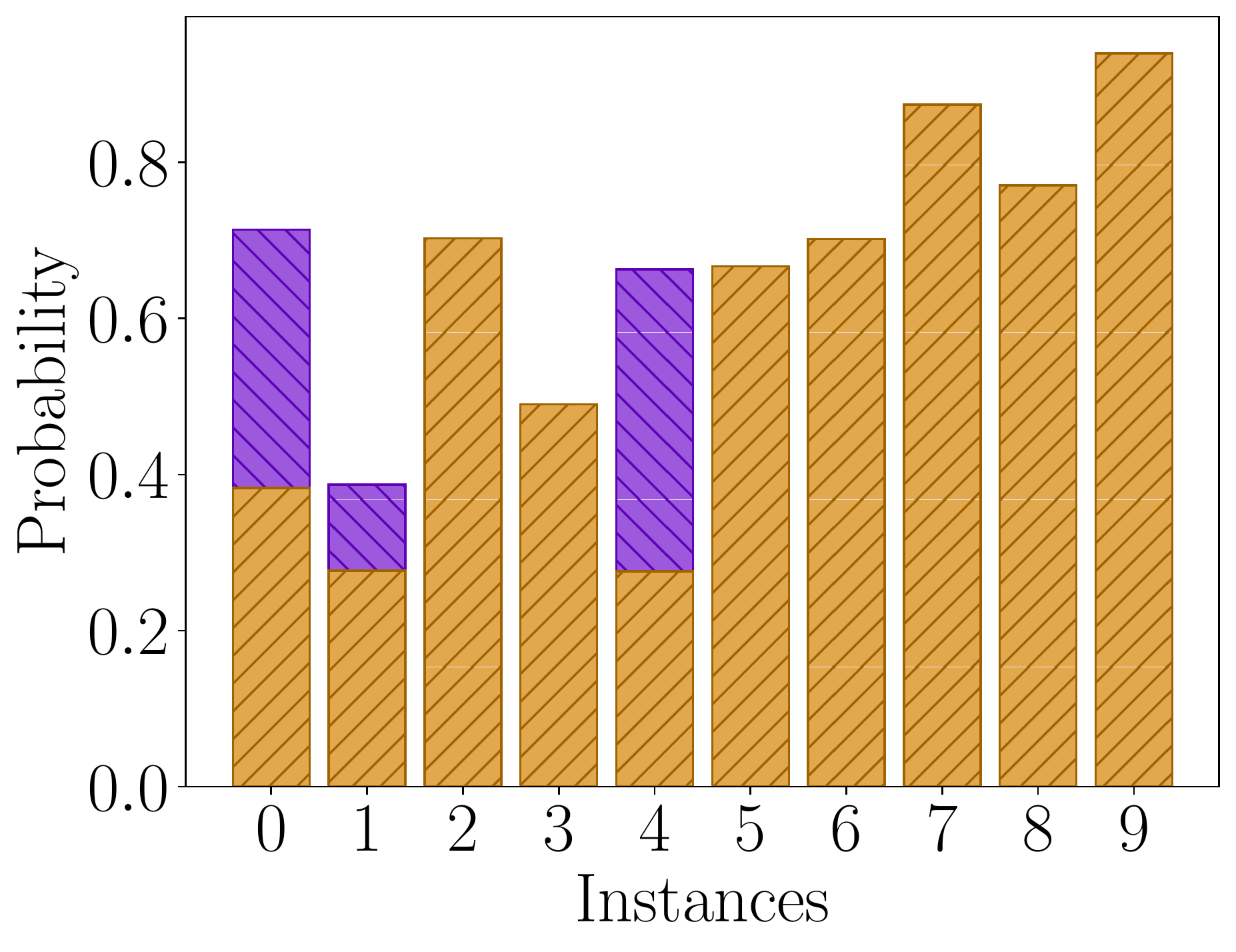}
		\caption{}
	\end{subfigure}
	\begin{subfigure}[b]{0.30\linewidth}
		\centering
		\includegraphics[height=3.5cm]{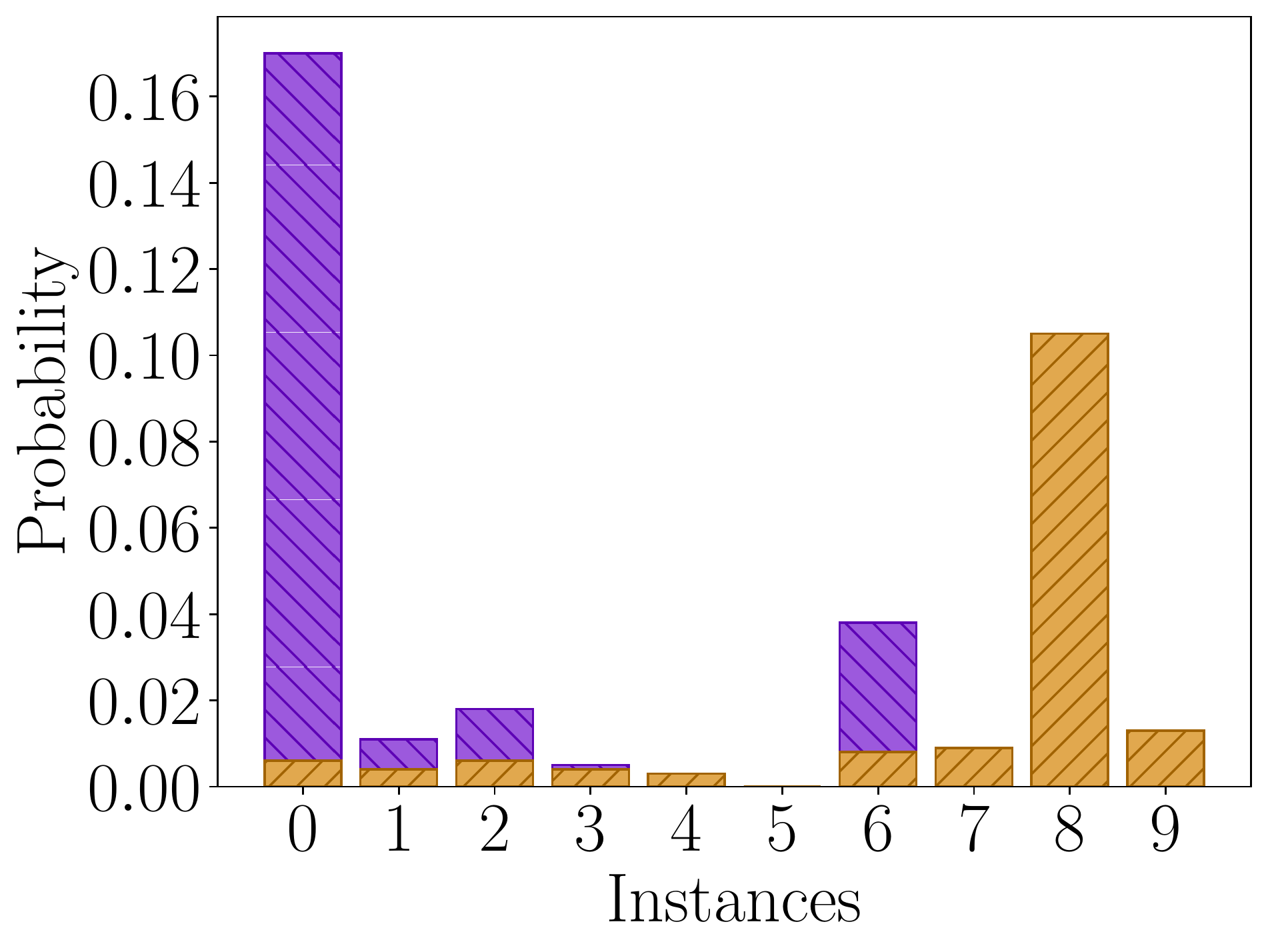} 
			\caption{}
	\end{subfigure} 
	\begin{subfigure}[b]{0.30\linewidth}
		\centering
		\includegraphics[width=1\linewidth]{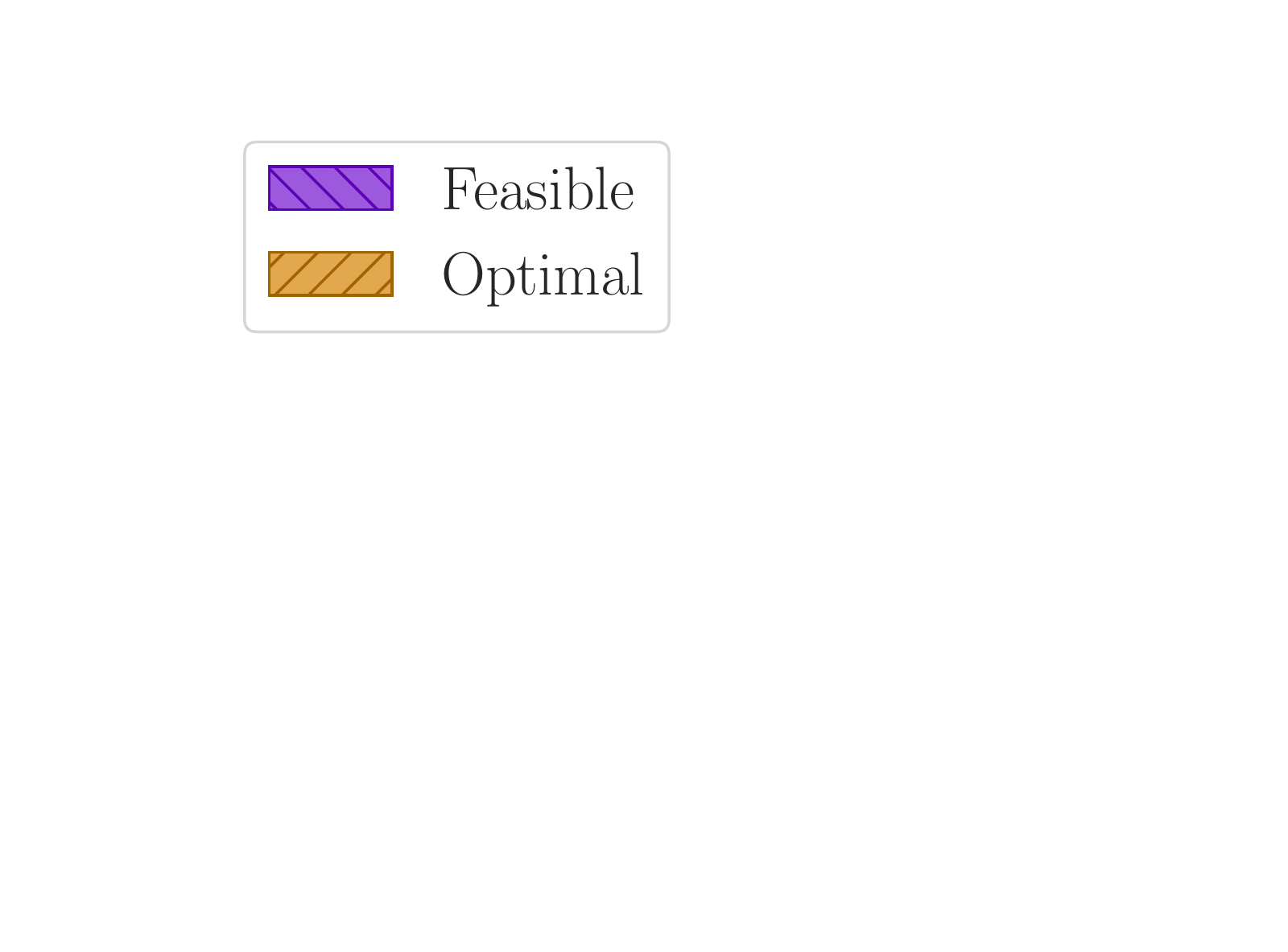} 
	\end{subfigure} \hspace{1in}
	\begin{subfigure}[b]{0.30\linewidth}
		\centering
		\includegraphics[height=3.5cm]{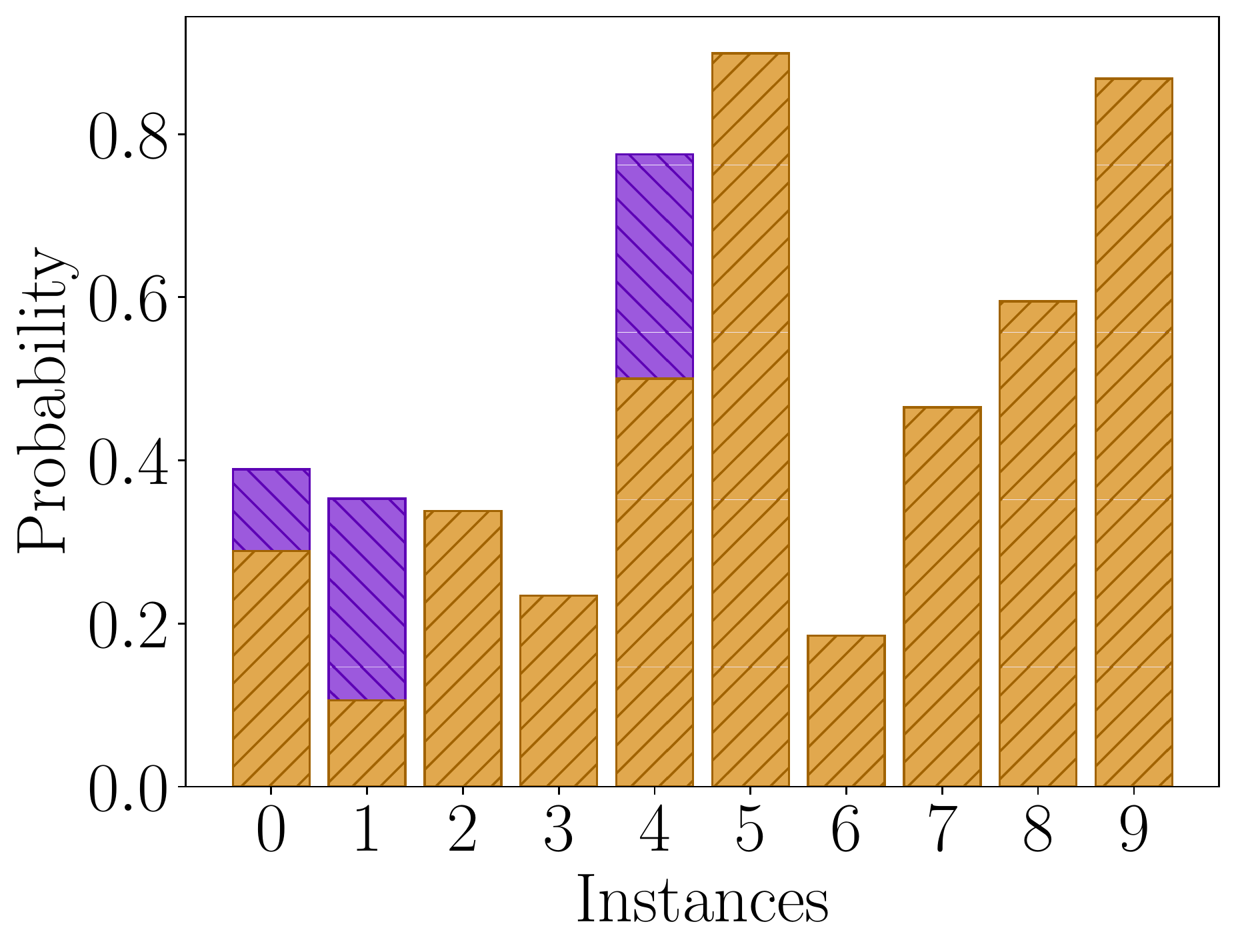}
				\caption{}
	\end{subfigure}
	\begin{subfigure}[b]{0.30\linewidth}
		\centering
		\includegraphics[height=3.5cm]{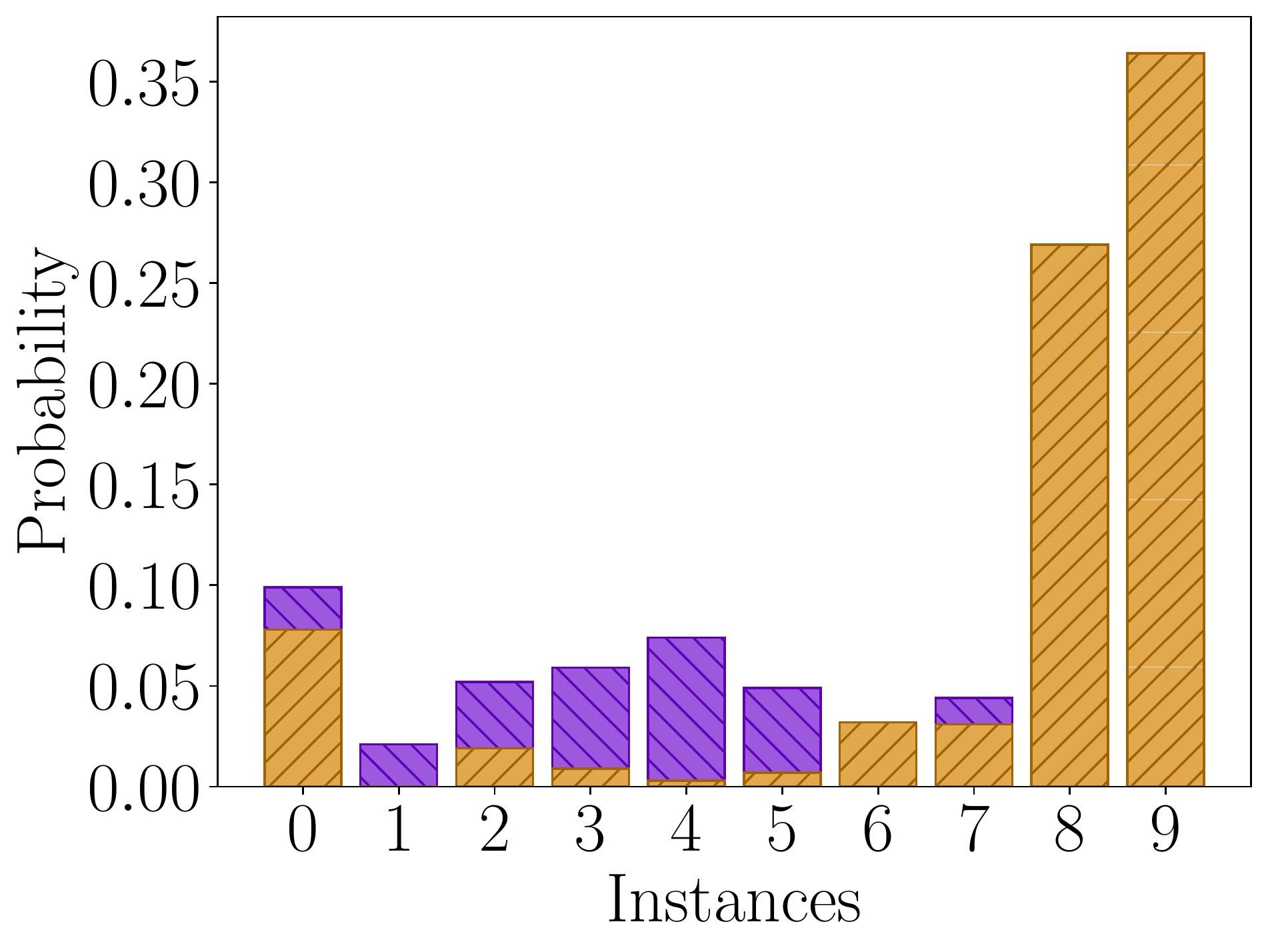}
				\caption{}
	\end{subfigure}
	\begin{subfigure}[b]{0.30\linewidth}
		\centering
		\includegraphics[height=3.5cm]{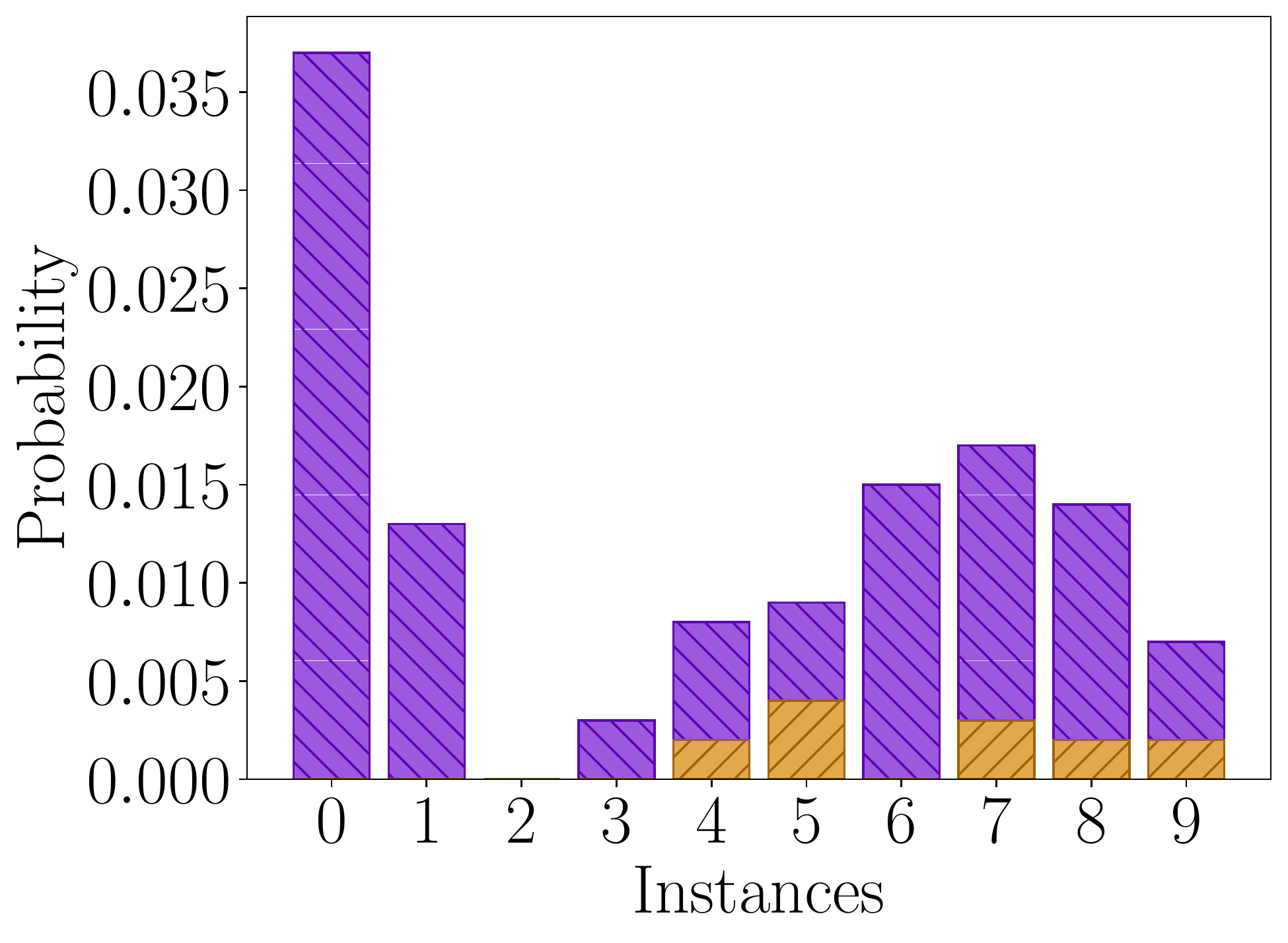}
				\caption{}
	\end{subfigure}
	\caption{Ratio of samples that encode feasible and optimal routes is given in the following plots. a) Edge-based formulation for instances with 3 cities. b) Edge-based formulation for instances with 4 cities. The sample with the lowest energy does not encode optimal route for instances numbered 1, 5, and 6. c) ILP formulation for instances with 3 cities. d) ILP formulation for instances with 4 cities. The sample with the lowest energy does not encode the optimal route for instances numbered 2, 3, 4, 5, and 7. e) ILP formulation for instances with 5 cities. The lowest energy sample does not encode the optimal route for any of the instances.  }
	\label{fig:probs}
\end{figure}

Besides checking whether the lowest energy state encodes the optimal route, we calculated the ratio of the samples encoding feasible and optimal routes among the sampleset, which is displayed in Figure \ref{fig:probs}. For the instances with 5 cities, no samples encoding feasible routes and optimal routes were observed for the edge-based formulation.

%%%%%%%%%%%%%%%%%%%%%%%%%%%%%%%%%%%%%%%%%%%%%%%%%%%%%%%%%%%%%%%%%%%%%%%%%%%%%%%%
\section{Discussion} \label{sec:discussion}
%%%%%%%%%%%%%%%%%%%%%%%%%%%%%%%%%%%%%%%%%%%%%%%%%%%%%%%%%%%%%%%%%%%%%%%%%%%%%%%%

Let us now analyze some key features of the formulations presented in this work and discuss the experimental results obtained from the D-Wave machine.

When we compare the number of qubits required by each formulation, the node-based formulation is the most advantageous when $|\edges| \gg n$. A similar observation can be also made by looking at Table~\ref{tbl: vars} in the Appendix, which compares the number of variables required for the instances from the AFG dataset. On the other hand, one of the main challenges in mapping real world-applications to current D-Wave devices is the restriction of 2-local interactions. One often ends up with a higher-order problem, as in the case of our node-based formulation, and some quadratization is required to recast the problem into QUBO format. If one applies quadratization to the node-based model using the procedure described in \cite{rosenberg1975reduction}, the resulting QUBO model has asymptotically the same number of variables as the original QUBO model. More information regarding quadratization is given in the Appendix. However, the inclusion of additional constraints due to quadratization introduces new constants to be tuned. Adjustment of these parameters is one of the challenges in QA in general.

In \cite{perdomo2019readiness}, the authors directly simulate HOBO models using simulated annealing and simulated quantum annealing (SQA) to assess whether there is any advantage over quadratization. It turns out that using the HOBO formulation does not improve the performance of the SQA which may be due to the limited connectivity of current devices. Nevertheless, future quantum annealers might allow $ k $-local interactions as well for $ k \geq 3 $ or better connectivity, which makes HOBO formulations still valuable. Furthermore, HOBO formulations can be natively solved using some other approaches like QAOA or VQE as suggested in \cite{tabi2020quantum,glos2020space,hadfield2019quantum,perdomo2019readiness,anschuetz2019variational}.

An advantage of the presented edge-based and node-based formulations for TSPTW is that they allow the vehicle to wait in a city even after the earliest start time. Hence the same route exists in the search space with different assignments to qubits representing waiting times and slack variables. Thanks to this, different bit assignments encode the optimal route without any penalty.

Recall that the main difference between the formulations is how we encode the binary variables, either using edge or node formulation. This dichotomy often appears when encoding problems related to graphs and it is also mentioned in \cite{kochenberger2014unconstrained}. Let us mention that there are alternative encoding ideas not covered in this paper. For instance, in \cite{glos2020space,tabi2020quantum}, the authors used binary encoding to represent the permutations for the TSP problem. Even though such encoding has an unbounded-order Pauli terms, an optimal number of $\sim n\log n$ qubits  was achieved. 

One of the biggest challenges for all formulations is the choice of penalty values. In the original TSP problem, the only constants to be adjusted are $ P_1 $ and $ P_3 $ and it is enough to set $ P_1 = 2 P_3 \max_{v,w}c_{v,w}$~\cite{lucas2014ising}.  In the case of TSPTW, this requires a more detailed investigation. In this study, we used simulated annealing to search for the penalty values, and the search was restricted to a predefined range. The penalty values heavily depend on the specific instance, making it hard to find a general rule which would work for all instances. Besides the penalty values, there are multiple parameters such as the annealing time or the chain strength that should be tuned while performing the experiments.  

For the instances we have used, the edge-based formulation requires fewer qubits than the ILP formulation. However, for large instances, ILP formulation might be more advantageous. Furthermore, when the experimental results for the edge-based and ILP formulations are compared, it is observed that ILP formulation is more promising as the probability of observing a sample encoding the optimal route is higher for the instances with 4 cities, and some samples encoding feasible routes are observed for the instances with 5 cities. The difference between the performance of the two formulations might be due to the different energy landscapes created by the two formulations. Another reason might be the choice of correct penalty constants. Overall, long chain lengths are one of the reasons behind the unsatisfactory results.

%%%%%%%%%%%%%%%%%%%%%%%%%%%%%%%%%%%%%%%%%%%%%%%%%%%%%%%%%%%%%%%%%%%%%%%%%%%%%%%%
\section{Conclusions and Future Work}\label{sec:conclusion}
%%%%%%%%%%%%%%%%%%%%%%%%%%%%%%%%%%%%%%%%%%%%%%%%%%%%%%%%%%%%%%%%%%%%%%%%%%%%%%%%

In this paper we proposed three unconstrained binary models for general  Travelling Salesman Problem with Time windows. Two of the introduced models were QUBO models, which (up to graph embedding) can be natively used for quantum annealing. The third model includes higher-order terms which makes it more suitable to variational quantum computing or digital annealers. We investigated analytically the memory requirements of the introduced models. Finally, we investigated the performance of edge-based model and ILP model on the currently available quantum annealer developed by D-Wave, using randomly created TSP instances with 3, 4 and 5 cities. ILP model performed better as the the probability of observing a sample encoding the optimal route was higher for the 4 cities case. In addition, some samples encoding feasible routes were observed in the case of 5 cities in the ILP model, while no feasible samples were obtained in the edge-based model.

A natural progression of this work is to investigate further the choice of penalty values. Instead of assigning a single penalty value for all of the time windows constraints, using more than one penalty value for different sub-constraints can help fine-tuning. For the node-based formulation, some other techniques for quadratization \cite{dattani2019quadratization, mandal2020compressed} can be investigated, which can result in a model with less resource requirement. Further studies should be carried out to examine more closely the energy distribution of the samples, and in particular the gap between the energy of the ground state and the samples obtained from D-Wave.  

The formulations presented here may be extended to a wide range of problems based on TSP like the vehicle routing problem and its variants, laying the groundwork for future research into the field of quantum optimization. Although the experimental work is limited to small instances, it offers insights into the field, emphasizing some of the challenges faced while solving complicated optimization problems using QA. Considering the number of variables required by the real instances, the study also puts forwards the limits of the current quantum hardware for solving real-world problems suggesting that hybrid algorithms can be a better alternative for solving large instances. 

On should also note that alternative proposals for harnessing binary models have been proposed. In particular, in \cite{tsukamoto2017accelerator} quantum-inspired hardware architecture for speeding-up solutions of combinatorial problems was introduced. The architecture has been implemented using FPGA, enabling over four orders of magnitude of speedup for solving TSP comparing with simulated annealing. From the perspective of real-world applications, TSPTW provides more challenging and more relevant problem to study.

%%%%%%%%%%%%%%%%%%%%%%%%%%%%%%%%%%%%%%%%%%%%%%%%%%%%%%%%%%%%%%%%%%%%%%%%%%%%%%%%
\paragraph{Acknowledgement}
This work has been partially supported by Polish National Science Center under the grant agreement 2019/33/B/ST6/02011. AG has been also supported by Polish National Science Center under the grant agreement 2020/37/N/ST6/02220. We would like to thank İmdat Kara for discussion on their results on \cite{kara2013new}.

%%%%%%%%%%%%%%%%%%%%%%%%%%%%%%%%%%%%%%%%%%%%%%%%%%%%%%%%%%%%%%%%%%%%%%%%%%%%%%%%
\paragraph{Data Availability}
The code used for generating the data is available on \url{https://doi.org/10.5281/zenodo.4966588}.
\bibliographystyle{ieeetr}
\bibliography{tsptw}

\begin{thebibliography}{10}

\bibitem{tsp2002}
G.~Gutin and A.~Punnen, eds., {\em The Traveling Salesman Problem and its
  Variations}.
\newblock Combinatorial Optimization, Kluwer Academic Press, 2002.

\bibitem{tsp2011}
V.~Chvatal, D.~L. Applegate, R.~E. Bixby, and W.~J. Cook, {\em The Traveling
  Salesman Problem: A Computational Study}.
\newblock Princeton Series in Applied Mathematics, Princeton University Press,
  2011.

\bibitem{desrosiers1995time}
J.~Desrosiers, Y.~Dumas, M.~M. Solomon, and F.~Soumis, ``Time constrained
  routing and scheduling,'' {\em Handbooks in operations research and
  management science}, vol.~8, pp.~35--139, 1995.

\bibitem{el2010vehicle}
N.~A. El-Sherbeny, ``Vehicle routing with time windows: An overview of exact,
  heuristic and metaheuristic methods,'' {\em Journal of King Saud
  University-Science}, vol.~22, no.~3, pp.~123--131, 2010.

\bibitem{savelsbergh1985local}
M.~W. Savelsbergh, ``Local search in routing problems with time windows,'' {\em
  Annals of Operations research}, vol.~4, no.~1, pp.~285--305, 1985.

\bibitem{preskill2018quantum}
J.~Preskill, ``{Quantum Computing} in the {NISQ} era and beyond,'' {\em
  Quantum}, vol.~2, p.~79, 2018.

\bibitem{peruzzo2014variational}
A.~Peruzzo, J.~McClean, P.~Shadbolt, M.-H. Yung, X.-Q. Zhou, P.~J. Love,
  A.~Aspuru-Guzik, and J.~L. O’brien, ``A variational eigenvalue solver on a
  photonic quantum processor,'' {\em Nature Communications}, vol.~5, no.~1,
  pp.~1--7, 2014.

\bibitem{farhi2014quantum}
E.~Farhi, J.~Goldstone, and S.~Gutmann, ``A quantum approximate optimization
  algorithm,'' {\em arXiv preprint arXiv:1411.4028}, 2014.

\bibitem{apolloni1989quantum}
B.~Apolloni, C.~Carvalho, and D.~De~Falco, ``Quantum stochastic optimization,''
  {\em Stochastic Processes and their Applications}, vol.~33, no.~2,
  pp.~233--244, 1989.

\bibitem{kadowaki1998quantum}
T.~Kadowaki and H.~Nishimori, ``Quantum annealing in the transverse {I}sing
  model,'' {\em Physical Review E}, vol.~58, no.~5, p.~5355, 1998.

\bibitem{farhi2000quantum}
E.~Farhi, J.~Goldstone, S.~Gutmann, and M.~Sipser, ``Quantum computation by
  adiabatic evolution,'' {\em arXiv preprint quant-ph/0001106}, 2000.

\bibitem{childs2001robustness}
A.~M. Childs, E.~Farhi, and J.~Preskill, ``Robustness of adiabatic quantum
  computation,'' {\em Physical Review A}, vol.~65, no.~1, p.~012322, 2001.

\bibitem{McGeoch2020theory}
C.~C. McGeoch, ``Theory versus practice in annealing-based quantum computing,''
  {\em Theoretical Computer Science}, vol.~816, pp.~169--183, 2020.

\bibitem{hauke2020perspectives}
P.~Hauke, H.~G. Katzgraber, W.~Lechner, H.~Nishimori, and W.~D. Oliver,
  ``Perspectives of quantum annealing: Methods and implementations,'' {\em
  Reports on Progress in Physics}, vol.~83, no.~5, p.~054401, 2020.

\bibitem{ronnow2014defining}
T.~F. R{\o}nnow, Z.~Wang, J.~Job, S.~Boixo, S.~V. Isakov, D.~Wecker, J.~M.
  Martinis, D.~A. Lidar, and M.~Troyer, ``Defining and detecting quantum
  speedup,'' {\em science}, vol.~345, no.~6195, pp.~420--424, 2014.

\bibitem{johnson2011quantum}
M.~W. Johnson, M.~H. Amin, S.~Gildert, T.~Lanting, F.~Hamze, N.~Dickson,
  R.~Harris, A.~J. Berkley, J.~Johansson, P.~Bunyk, {\em et~al.}, ``Quantum
  annealing with manufactured spins,'' {\em Nature}, vol.~473, no.~7346,
  pp.~194--198, 2011.

\bibitem{papalitsas2019qubo}
C.~Papalitsas, T.~Andronikos, K.~Giannakis, G.~Theocharopoulou, and
  S.~Fanarioti, ``A {QUBO} model for the traveling salesman problem with time
  windows,'' {\em Algorithms}, vol.~12, no.~11, p.~224, 2019.

\bibitem{kara2013new}
I.~Kara, O.~N. Koc, F.~Alt{\i}parmak, and B.~Dengiz, ``New integer linear
  programming formulation for the traveling salesman problem with time windows:
  Minimizing tour duration with waiting times,'' {\em Optimization}, vol.~62,
  no.~10, pp.~1309--1319, 2013.

\bibitem{kirkpatrick1983optimization}
S.~Kirkpatrick, C.~D. Gelatt, and M.~P. Vecchi, ``Optimization by simulated
  annealing,'' {\em science}, vol.~220, no.~4598, pp.~671--680, 1983.

\bibitem{mcgeoch2014adiabatic}
C.~C. McGeoch, ``Adiabatic quantum computation and quantum annealing: Theory
  and practice,'' {\em Synthesis Lectures on Quantum Computing}, vol.~5, no.~2,
  pp.~1--93, 2014.

\bibitem{lucas2014ising}
A.~Lucas, ``Ising formulations of many {NP} problems,'' {\em Frontiers in
  Physics}, vol.~2, p.~5, 2014.

\bibitem{perdomo2019readiness}
A.~Perdomo-Ortiz, A.~Feldman, A.~Ozaeta, S.~V. Isakov, Z.~Zhu, B.~O’Gorman,
  H.~G. Katzgraber, A.~Diedrich, H.~Neven, J.~de~Kleer, {\em et~al.},
  ``Readiness of quantum optimization machines for industrial applications,''
  {\em Physical Review Applied}, vol.~12, no.~1, p.~014004, 2019.

\bibitem{baker1983exact}
E.~K. Baker, ``An exact algorithm for the time-constrained travelling salesman
  problem,'' {\em Operations Research}, vol.~31, no.~5, pp.~938--945, 1983.

\bibitem{langevin1993two}
A.~Langevin, M.~Desrochers, J.~Desrosiers, S.~G{\'e}linas, and F.~Soumis, ``A
  two-commodity flow formulation for the traveling salesman and the makespan
  problems with time windows,'' {\em Networks}, vol.~23, no.~7, pp.~631--640,
  1993.

\bibitem{baldacci2012new}
R.~Baldacci, A.~Mingozzi, and R.~Roberti, ``New state-space relaxations for
  solving the traveling salesman problem with time windows,'' {\em INFORMS
  Journal on Computing}, vol.~24, no.~3, pp.~356--371, 2012.

\bibitem{christofides1981state}
N.~Christofides, A.~Mingozzi, and P.~Toth, ``State-space relaxation procedures
  for the computation of bounds to routing problems,'' {\em Networks}, vol.~11,
  no.~2, pp.~145--164, 1981.

\bibitem{dumas1995optimal}
Y.~Dumas, J.~Desrosiers, E.~Gelinas, and M.~M. Solomon, ``An optimal algorithm
  for the travelling salesman problem with time windows,'' {\em Operations
  Research}, vol.~43, no.~2, pp.~367--371, 1995.

\bibitem{pesant1998exact}
G.~Pesant, M.~Gendreau, J.-Y. Potvin, and J.-M. Rousseau, ``An exact constraint
  logic programming algorithm for the traveling salesman problem with time
  windows,'' {\em Transportation Science}, vol.~32, no.~1, pp.~12--29, 1998.

\bibitem{focacci2002hybrid}
F.~Focacci, A.~Lodi, and M.~Milano, ``A hybrid exact algorithm for the
  {TSPTW},'' {\em INFORMS Journal on Computing}, vol.~14, no.~4, pp.~403--417,
  2002.

\bibitem{cappart2020combining}
Q.~Cappart, T.~Moisan, L.-M. Rousseau, I.~Pr{\'e}mont-Schwarz, and A.~Cire,
  ``Combining reinforcement learning and constraint programming for
  combinatorial optimization,'' {\em arXiv preprint arXiv:2006.01610}, 2020.

\bibitem{hadfield2017quantum}
S.~Hadfield, Z.~Wang, E.~G. Rieffel, B.~O'Gorman, D.~Venturelli, and R.~Biswas,
  ``Quantum approximate optimization with hard and soft constraints,'' in {\em
  Proceedings of the Second International Workshop on Post Moores Era
  Supercomputing}, pp.~15--21, 2017.

\bibitem{hadfield2019quantum}
S.~Hadfield, Z.~Wang, B.~O’Gorman, E.~G. Rieffel, D.~Venturelli, and
  R.~Biswas, ``From the quantum approximate optimization algorithm to a quantum
  alternating operator ansatz,'' {\em Algorithms}, vol.~12, no.~2, p.~34, 2019.

\bibitem{glos2020space}
A.~Glos, A.~Krawiec, and Z.~Zimbor{\'a}s, ``Space-efficient binary optimization
  for variational computing,'' {\em arXiv preprint arXiv:2009.07309}, 2020.

\bibitem{martovnak2004quantum}
R.~Marto{\v{n}}{\'a}k, G.~E. Santoro, and E.~Tosatti, ``Quantum annealing of
  the travelling-salesman problem,'' {\em Physical Review E}, vol.~70, no.~5,
  p.~057701, 2004.

\bibitem{santoro2006optimization}
G.~E. Santoro and E.~Tosatti, ``Optimization using quantum mechanics: quantum
  annealing through adiabatic evolution,'' {\em Journal of Physics A:
  Mathematical and General}, vol.~39, no.~36, p.~R393, 2006.

\bibitem{borowski2020new}
M.~Borowski, P.~Gora, K.~Karnas, M.~B{\l}ajda, K.~Kr{\'o}l, A.~Matyjasek,
  D.~Burczyk, M.~Szewczyk, and M.~Kutwin, ``New hybrid quantum annealing
  algorithms for solving vehicle routing problem,'' in {\em International
  Conference on Computational Science}, pp.~546--561, Springer, 2020.

\bibitem{irie2019quantum}
H.~Irie, G.~Wongpaisarnsin, M.~Terabe, A.~Miki, and S.~Taguchi, ``Quantum
  annealing of vehicle routing problem with time, state and capacity,'' in {\em
  International Workshop on Quantum Technology and Optimization Problems},
  pp.~145--156, Springer, 2019.

\bibitem{ascheuer1996hamiltonian}
N.~Ascheuer, ``Hamiltonian path problems in the on-line optimization of
  flexible manufacturing systems,'' 1996.

\bibitem{boothby2020next}
K.~Boothby, P.~Bunyk, J.~Raymond, and A.~Roy, ``Next-generation topology of
  {D-Wave} quantum processors,'' {\em arXiv preprint arXiv:2003.00133}, 2020.

\bibitem{rosenberg1975reduction}
I.~G. Rosenberg, ``Reduction of bivalent maximization to the quadratic case,''
  1975.

\bibitem{tabi2020quantum}
Z.~Tabi, K.~H. El-Safty, Z.~Kallus, P.~H{\'a}ga, T.~Kozsik, A.~Glos, and
  Z.~Zimbor{\'a}s, ``Quantum optimization for the graph coloring problem with
  space-efficient embedding,'' in {\em 2020 IEEE International Conference on
  Quantum Computing and Engineering (QCE)}, pp.~56--62, IEEE, 2020.

\bibitem{anschuetz2019variational}
E.~Anschuetz, J.~Olson, A.~Aspuru-Guzik, and Y.~Cao, ``Variational quantum
  factoring,'' in {\em International Workshop on Quantum Technology and
  Optimization Problems}, pp.~74--85, Springer, 2019.

\bibitem{kochenberger2014unconstrained}
G.~Kochenberger, J.-K. Hao, F.~Glover, M.~Lewis, Z.~L{\"u}, H.~Wang, and
  Y.~Wang, ``The unconstrained binary quadratic programming problem: a
  survey,'' {\em Journal of Combinatorial Optimization}, vol.~28, no.~1,
  pp.~58--81, 2014.

\bibitem{dattani2019quadratization}
N.~Dattani, ``Quadratization in discrete optimization and quantum mechanics,''
  {\em arXiv preprint arXiv:1901.04405}, 2019.

\bibitem{mandal2020compressed}
A.~Mandal, A.~Roy, S.~Upadhyay, and H.~Ushijima-Mwesigwa, ``Compressed
  quadratization of higher order binary optimization problems,'' in {\em
  Proceedings of the 17th ACM International Conference on Computing Frontiers},
  pp.~126--131, 2020.

\bibitem{tsukamoto2017accelerator}
S.~Tsukamoto, M.~Takatsu, S.~Matsubara, and H.~Tamura, ``An accelerator
  architecture for combinatorial optimization problems,'' {\em Fujitsu Sci.
  Tech. J}, vol.~53, no.~5, pp.~8--13, 2017.

\end{thebibliography}

%\newpage 
\appendix
%%%%%%%%%%%%%%%%%%%%%%%%%%%%%%%%%%%%%%%%%%%%%%%%%%%%%%%%%%%%%%%%%%%%%%%%%%%%%%%%
\section{Quadratization} 
%%%%%%%%%%%%%%%%%%%%%%%%%%%%%%%%%%%%%%%%%%%%%%%%%%%%%%%%%%%%%%%%%%%%%%%%%%%%%%%%

Given a pseudo-Boolean function $ f(x) $ on $ \{0,1\}^n $, $ g(x,y) $ is a quadratization of $ f $ if 
\[ 
f(x) = \min \{g(x,y): y \in \{0,1\}^m\} \mbox{ for all } x \in \{0,1\}^n,
\] 
where $ y $ is a set of auxiliary variables $ y_1,y_2,\dots,y_m $. Using quadratization, one can reduce the problem of minimizing HOBO into the problem of minimizing QUBO. One of the proposed procedures for quadratization is given by Rosenberg \cite{rosenberg1975reduction} and goes as follows:
\begin{itemize}
	\item Pick two variables $ x_i$,  $x_j $ such that the product $ x_ix_j $ appears in a term with degree at least 3.
	\item Replace each occurrence of $ x_ix_j $ with a new variable $ y_{ij} \in \{0,1\} $.
	\item Add the penalty term $C(x_ix_j - 2 x_i y_{ij} - 2x_j y_{ij} + 3 y_{ij}) $ where $ C $ is a constant.	
\end{itemize}

In our case, the higher order terms in the node-based formulation appears from the time windows constraints, in particular from the expression for the arrival times:
\[ \tilde{A}_i = \sum_{t=1}^{i-1} W_t + \tilde{A}_1 + \sum_{t=2}^{i} \sum_{\substack{u,v=1\\u \neq v}}^n  c_{uv}x_u^{t-1}x_v^{t}. \]
The square of $ A_i $ appears in the Hamiltonian, resulting in 3-local and 4-local terms involving the product $x_u^{t-1}x_v^{t} $. By replacing each occurrence of $x_u^{t-1}x_v^{t} $ for $ t=2,\dots, n $ and $ u,v \in \{1,\dots,n\} $ by a new variable $x_{u,v}^{t} $, we may reduce or model to QUBO, at the cost of additional $ (n-1)n(n-1) $ variables and penalty terms. 

After quadratization, the number of variables (excluding the variables representing the waiting times and the slack variables) required by the node-based model is $ n^3-n^2+n = O(n^3) $, which is asymptotically equal to the number of variables required by the edge-based formulation.

%%%%%%%%%%%%%%%%%%%%%%%%%%%%%%%%%%%%%%%%%%%%%%%%%%%%%%%%%%%%%%%%%%%%%%%%%%%%%%%%
\section{Resource analysis for real instances}
%%%%%%%%%%%%%%%%%%%%%%%%%%%%%%%%%%%%%%%%%%%%%%%%%%%%%%%%%%%%%%%%%%%%%%%%%%%%%%%%
In Table \ref{tbl: vars}, we present the number of variables required by instances from the AFG dataset introduced by Ascheuer in \cite{ascheuer1996hamiltonian}. The instances represent real world problems of optimization of manufacturing systems. We point out the number of variables needed for each formulation and the percentage of the variables that encode the slack variables.

% \usepackage{booktabs}
\begin{table}[]

	\caption{Analysis for the number of variables for the AFG dataset.}
		\label{tbl: vars}
	\centering
	\tiny
	\begin{tabular}{@{}cccccccc@{}}
		\toprule
		\textbf{Instance name} & \textbf{Cities} & \textbf{QUBO} & \textbf{Slack \%} & \textbf{HOBO} & \textbf{Slack \%} & \textbf{ILP} & \textbf{Slack \%} \\ \midrule
		rbg016a.            & 17              & 2830          & 14.7              & 864           & 48.1              & 4103         & 88.7              \\ \midrule
		rbg016b.            & 17              & 3414          & 12.2              & 848           & 49.1              & 5353         & 90.2              \\ \midrule
		rbg017.2.           & 16              & 3190          & 12.2              & 795           & 49.1              & 5405         & 90.2              \\ \midrule
		rbg017.             & 16              & 2602          & 15                & 795           & 49.1              & 4220         & 88.8              \\ \midrule
		rbg017a.            & 18              & 4634          & 10.3              & 969           & 49.1              & 7382         & 91.3              \\ \midrule
		rbg019a.            & 20              & 4144          & 11.9              & 1083          & 45.6              & 5358         & 89.1              \\ \midrule
		rbg019b.            & 20              & 5334          & 10                & 1121          & 47.5              & 7226         & 90.7              \\ \midrule
		rbg019c.            & 20              & 6199          & 8.6               & 1140          & 46.7              & 8962         & 91.8              \\ \midrule
		rbg019d.            & 20              & 4666          & 10.6              & 1083          & 45.6              & 6335         & 89.9              \\ \midrule
		rbg020a.            & 21              & 6750          & 8.3               & 1220          & 45.9              & 9331         & 91.7              \\ \midrule
		rbg021.2.           & 20              & 6289          & 8.5               & 1140          & 46.7              & 9152         & 91.8              \\ \midrule
		rbg021.3.           & 20              & 6307          & 8.4               & 1140          & 46.7              & 9231         & 91.9              \\ \midrule
		rbg021.4.           & 20              & 6397          & 8.3               & 1140          & 46.7              & 9505         & 92                \\ \midrule
		rbg021.5.           & 20              & 6433          & 8.3               & 1140          & 46.7              & 9687         & 92.1              \\ \midrule
		rbg021.6.           & 20              & 6757          & 7.9               & 1140          & 46.7              & 10296        & 92.3              \\ \midrule
		rbg021.7.           & 20              & 6975          & 8.2               & 1178          & 48.4              & 10837        & 92.4              \\ \midrule
		rbg021.8.           & 20              & 7011          & 8.1               & 1178          & 48.4              & 10903        & 92.5              \\ \midrule
		rbg021.9.           & 20              & 7011          & 8.1               & 1178          & 48.4              & 10907        & 92.5              \\ \midrule
		rbg021.             & 20              & 6199          & 8.6               & 1140          & 46.7              & 8962         & 91.8              \\ \midrule
		rbg027a.            & 28              & 16787         & 4.5               & 1836          & 41.2              & 17627        & 92.9              \\ \midrule
		rbg031a.            & 32              & 19030         & 4.2               & 2139          & 37.7              & 16108        & 92                \\ \midrule
		rbg033a.            & 34              & 22538         & 4.1               & 2409          & 38.4              & 18134        & 92.3              \\ \midrule
		rbg034a.            & 35              & 25615         & 3.5               & 2448          & 36.1              & 20143        & 92.5              \\ \midrule
		rbg035a.2.          & 36              & 36088         & 3.3               & 2835          & 42                & 33985        & 94.8              \\ \midrule
		rbg035a.            & 36              & 27002         & 3.4               & 2555          & 35.6              & 20549        & 92.5              \\ \midrule
		rbg038a.            & 39              & 33972         & 3.1               & 3002          & 35.4              & 25374        & 93                \\ \midrule
		rbg040a.            & 41              & 39121         & 2.9               & 3240          & 34.6              & 27680        & 93.1              \\ \midrule
		rbg041a.            & 42              & 42722         & 2.7               & 3321          & 34.6              & 27999        & 93                \\ \midrule
		rbg042a.            & 43              & 48010         & 2.3               & 3360          & 32.5              & 29996        & 93.1              \\ \midrule
		rbg048a.            & 49              & 86761         & 1.7               & 4416          & 32.6              & 54787        & 94.5              \\ \midrule
		rbg049a.            & 50              & 82318         & 1.8               & 4557          & 32.3              & 52603        & 94.4              \\ \midrule
		rbg050a.            & 51              & 98585         & 1.5               & 4700          & 31.9              & 60852        & 94.6              \\ \midrule
		rbg050b.            & 51              & 88932         & 1.7               & 4700          & 31.9              & 55113        & 94.4              \\ \midrule
		rbg050c.            & 51              & 96282         & 1.6               & 4700          & 31.9              & 59109        & 94.6              \\ \midrule
		rbg055a.            & 56              & 95569         & 1.6               & 5280          & 29.2              & 48909        & 93.8              \\ \midrule
		rbg067a.            & 68              & 166366        & 1.2               & 7437          & 27                & 73776        & 94.4              \\ \midrule
		rbg086a.            & 87              & 341749        & 0.8               & 11438         & 24.1              & 121615       & 95                \\ \midrule
		rbg092a.            & 93              & 425653        & 0.6               & 12512         & 22.1              & 137271       & 95                \\ \midrule
		rbg125a.            & 126             & 1038546       & 0.4               & 21125         & 17.8              & 244122       & 95.3              \\ \midrule
		rbg132.2.           & 131             & 1220131       & 0.3               & 23010         & 18.1              & 286488       & 95.5              \\ \midrule
		&                 &               &                   &               &                   &              &                   \\ \bottomrule
	\end{tabular}
\end{table}
\end{document}